\documentclass[10pt]{article}

\usepackage[a4paper,bindingoffset=0.2in,%
            left=0.3 in,right=0.5 in,top=0.5 in,bottom=0.5  in,%
            footskip=.25in]{geometry}

\usepackage{nccmath}

{\end{pmatrix}\end{medsize}}%

\newcommand\beq{\begin{equation}}
\newcommand\eeq{\end{equation}}
\newcommand\bea{\begin{eqnarray}}
\newcommand\eea{\end{eqnarray}}

\usepackage{physics,amsmath,amssymb,graphicx,enumitem ,color,dsfont}
\usepackage{subfigure,wrapfig}

\title{Spectral distances on doubled Moyal plane using Dirac eigen-spinors}

\begin{large}
\author{Kaushlendra Kumar\footnote{Indian Institute of Science Education and Research, Kolkata 741246, India; Email: kk15ip019@iiserkol.ac.in},~ Biswajit Chakraborty\footnote{S.N.Bose National Centre For Basic Sciences, Salt Lake, Kolkata 700098, India; Email: biswajit@bose.res.in}}
\end{large}

\begin{document}

\date{}

\maketitle
\begin{abstract}

We present here a novel method of computing spectral distances in doubled Moyal plane in a noncommutative geometrical framework using Dirac eigen-spinors, while solving the Lipschitz ball condition explicitly through matrices. The standard results of longitudinal, transverse and hypotenuse distances between different pairs of pure states have been computed and Pythagorean equality between them have been re-produced. The issue of non-unital nature of Moyal plane algebra is taken care of through a sequence of projection operators constructed from Dirac eigen-spinors, which plays a crucial role throughout  this paper. At the end, a toy model of ``Higgs field" has been constructed by fluctuating the Dirac operator and the variation on the transverse distance has been demonstrated, through an explicit computation.

\end{abstract}

\section{Introduction}

Alain Connes through his non commutative geometry provided a new insight into the structure of Standard model of particle physics \cite{1st} (see \cite{b16} for review). Here essentially one captures the whole gauge symmetry of Standard model (Which is formulated using ``Almost commutative spacetime") through group of inner-automorphism of the algebra

\begin{equation} \label{alg}
\mathcal{A} = \mathds{C} \oplus \mathds{H} \oplus M_3 (\mathds{C}),
\end{equation}

One of the most remarkable feature of this formulation is that the Higgs field arises naturally here, along with other gauge fields of Yang Mills theory and provides a unified conceptual perspective regarding the geometrical origin. This formulation although quite successful in predicting or post-dicting almost  all the phenomenological observations made so far, including the computation of Higgs mass of $125 \, GeV$ \cite{Ali} is not equipped yet to address the issues of quantum gravity, which is expected to play a role in the vicinity of Plank energy scale $\sim 10^{19} \, GeV$. One rather expects the differentiable manifold $M$ representing (Euclidianized) spacetime to be replaced by a truly non commutative space as follows from some plausibility arguments due to Doplicher \emph{et. al.} \cite{Dop1}. The simplest of such model noncommutative spaces are 

\begin{equation} \label{quantum spaces}
\textrm{Moyal plane}~ (\mathds{R}^2_*): [\hat{x}_1, \hat{x}_2] = i \theta ~;~  \textrm{Fuzzy sphere}~ (\mathds{S}^2_*): [\hat{x}_i, \hat{x}_j] = i\lambda \epsilon_{ijk}\hat{x}_k.
\end{equation}

The computations of distances on a generic non commutative spaces has to be performed with the formalism of non commutative geometry formulated by Connes \cite{Connes1}. In this context the computations of spectral distance for which the algebra is given by \eqref{alg} can be shown to be equivalent to that of the triple where the algebra in \eqref{alg} can be replaced by just $\mathds{C}^2$ \cite{Martinetti3, b20}. This results from the fact that the incorporation of the experimental data on the Dirac operator renders the distance between any pair of states in $M_3(\mathds{C})$ to diverge and the algebra of quarternions $\mathds{H}$ has only one state \cite{Martinetti3}. Besides the algebra $\mathcal{A}=\mathds{C}^2$ was used by Suijlekom \emph{et. al.} \cite{b15} to discuss electrodynamics in a noncommutative geometric set-ups. With this internal space, described by the algebra $\mathcal{A}=\mathds{C}^2$ along with other data for the spectral triple, essentially duplicates the copy of the manifold $M$, or for that matter the ``noncommutative manifold" like $\mathds{R}^2_*$ or $\mathds{S}^2_*$.
           
The questions naturally arises about the computability of spectral distance between pair of states associated to the same or different manifold. This is an important question in it's own right, as it can uncover some geometrical features of Higgs field, as the presence or absence of the Higgs field can have a non-trivial impact on these distances. These questions were addressed already in the literature \cite{Martinetti3, b20, b15}. One is particularly interested to compute the distance between states belonging to one of the copies of algebra and its ``Clone" (the more precise meaning of this terminology will be explained in subsequent sections) belonging to the other algebra. We refer to such a distance as ``Transverse". On the other hand distance between pair of states belonging to same algebra to be referred as ``Longitudinal". Finally distance between any pair of pure states which are not clone of each others but belong to different algebra will be referred to as ``Hypotenuse"  distance. In \cite{Martinetti3} the authors have proved various theorems and laid down conditions for Pythagoras theorem to hold both for commutative manifold $M$, described by commutative manifold $C^*$-algebra and also extended them to noncommutative spaces like doubled Moyal plane \cite{b20}, which is necessarily described by non commutative $C^*$-algebra. Eventually, these analyses were also extended to study distances between non-pure states involving more general spectral triples to find that in the generic cases the Pyhtagoras equalitities are replaced by corresponding inequalities \cite{A} and analysed their relationship with Wasserstein distance $W$ of order 1, which occurs in the theory of optical transport \cite{villani}.
           
Here we would like to revisit the problem by explicitly constructing the Dirac eigen-spinors for the doubled Moyal plane, by making use of the corresponding eigen-spinors for single Moyal plane, introduced in \cite{b18} through a Hilbert-Schimdt operator formulation to describe non relativistic non commutative quantum mechanics \cite{b2}. This has the advantage that it involves completely an operator formulation and is quite transparent in execution as one has to deal with matrices only, albeit, big ones some times!. We are therefore forced to make use of $Mathematica$ and could reproduce all the existing results of the distances and eventually verify Pythagoras theorem. At the end, we have explicitly shown the variation of transverse distance in the presence of a prototype ``Higgs" field, which arises when the Dirac operator is fluctuated using a general one-form. 

The paper is organised as follows. In section \ref{sec2} we have set up the framework of our calculation, namely the Hilbert-Schmidt operatorial formulation using which the spectral triple for Moyal plane has been introduced. Moreover, a review of the computation of spectral distance on Moyal plane using Dirac eigen-spinors \cite{b18} along with the aspects of translational invariance on Moyal plane has been provided here. Then in section \ref{sec3}, the notion of doubling the spectral triple using two-point space has been introduced with Moyal plane, as an example as well as the idea of restricted spectral triple has been reviewed. The construction of eigen-spinors corresponding to the Dirac operator of doubled Moyal plane has been taken up in section \ref{sec4} which has then been used extensively in section \ref{section5} to compute distances in all the three cases viz. `transverse', `longitudinal' and `hypotenuse' one. Finally, in section \ref{sec6} the Dirac operator has been fluctuated using an one-form arising solely from a prototype ``Higgs" field and then the variation of transverse distance has been studied by restricting the spectral triple to that of two-point space again using a projection operator built out of the Dirac eigen-spinors.

\section{Review of Hilbert-Schmidt operator formalism and the spectral distances on Moyal plane $\left( \mathds{R}^2_{*} \right)$} \label{sec2}

The Hilbert space $\mathcal{H}_q$ furnishing a representation of the entire noncommutative Heisenberg algebra

\begin{equation} \label{a1}
\left[ \hat{X}_i , \hat{X}_j \right] = i\theta\epsilon_{ij} ~;~ \left[ \hat{X}_i , \hat{P}_j \right] = i\delta_{ij} ~;~ \left[ \hat{P}_i , \hat{P}_j \right] = 0
\end{equation}

for the Moyal plane is given by Hilbert-Schmidt operators $\psi (\hat{x}_1, \hat{x}_2) \in \mathcal{H}_q$, consisting of algebra elements generated by the polynomials of the position operators $(\hat{x}_1, \hat{x}_2)$, subject to the above coordinate sub-algebra i.e. the commutation relations $\left[\hat{x}_1, \hat{x}_2\right] = i\theta$ in \eqref{quantum spaces}, on which the above operators $\hat{X}_i$ and $\hat{P}_j$ act as,

\begin{equation} \label{a2}
\hat{X}_i \psi (\hat{x}_1, \hat{x}_2) = \hat{x}_i \psi (\hat{x}_1, \hat{x}_2)~,~ \hat{P}_i \psi (\hat{x}_1, \hat{x}_2) = \frac{1}{\theta}\epsilon_{ij} [\hat{x}_j, \psi (\hat{x}_1, \hat{x}_2)]
\end{equation}

These operators $\hat{x}_i$ and therefore $\psi (\hat{x}_1, \hat{x}_2)$, in turn, act on an auxiliary Hilbert space $\mathcal{H}_c$, which furnishes a representation of just this coordinate sub-algebra \eqref{quantum spaces} and is defined by

\begin{equation} \label{Hc}
\mathcal{H}_c = \text{Span} \left\lbrace  |n\rangle = \frac{\left( \hat{b}^\dagger \right)^n}{\sqrt{n!}} \ket{0} ~;~ \hat{b} \ket{0} = 0 ~,~ \hat{b} = \frac{\hat{x}_1 + i\hat{x}_2}{\sqrt{2\theta}} \right\rbrace _{n=0}^{\infty} ~;~ [\hat{b}, \hat{b}^\dagger] = 1
\end{equation}

Note that $\mathcal{H}_c$ is isomorphic to the Hilbert space of 1-D harmonic oscillator where the role of momentum is played by another spatial coordinate operator $\hat{x}_2$ and that of $\hbar$ by $\theta$. The inner product of $\mathcal{H}_q$ is defined through $\mathcal{H}_c$ as

\begin{equation} \label{a4}
\left( \psi_1 (\hat{x}_1, \hat{x}_2), \psi_2 (\hat{x}_1, \hat{x}_2) \right) = \Tr_{\mathcal{H}_c} \left( \psi^\dagger_1 (\hat{x}_1, \hat{x}_2), \psi_2 (\hat{x}_1, \hat{x}_2) \right) = \sum\limits_{n = 0}^{\infty} \bra{n} \psi^\dagger_1 (\hat{x}_1, \hat{x}_2), \psi_2 (\hat{x}_1, \hat{x}_2) \ket{n}
\end{equation}

This is clearly well-defined, as $\mathcal{H}_q$ is necessarily, by definition, comprises of elements which are compact and trace-class operators. Also note that we make a distinction here between $\hat{X}_i$ and $\hat{x}_i$, depending upon their domains of action i.e. on $\mathcal{H}_q$ and $\mathcal{H}_c$ respectively. In a sense, the former can be regarded as a representation of the latter. On the other hand, the momentum operator $\hat{P}_i$ has action on $\mathcal{H}_q$ only, as is clear from \eqref{a2}, where it is identified through the adjoint action of $\epsilon_{ij} \hat{x}_j$ and therefore corresponds to the difference between left and right actions on $\psi (\hat{x}_1, \hat{x}_2)$. Finally, the vectors of $\mathcal{H}_q$ and $\mathcal{H}_c$ are distinguished by using round $| . )$ and $\ket{.}$ kets respectively.

In view of noncommutative (i.e. $\theta \neq 0$) nature of the coordinate algebra common eigenstates of $\hat{x}_1$ and $\hat{x}_2$ cannot simply exist. One is therefore forced to introduce a coherent state

\begin{equation} \label{a5}
|z\rangle = e^{-\bar{z}\hat{b} + z\hat{b}^\dagger} \ket{0} = e^{-\abs{z}^2/2}e^{z\hat{b}^\dag}|0\rangle \in \mathcal{H}_c ~;~ \hat{b}\ket{z} = z \ket{z};~ z = \frac{x_1 + ix_2}{\sqrt{2\theta}}
\end{equation}

with maximal localization : $\Delta x_1 \Delta x_2 = \theta/2$. As is well-known, this provides an over-complete and non-orthonormal basis in $\mathcal{H}_c$ : $\bra{z^\prime}\ket{z} = e^{-\abs{z^\prime - z}^2/2}$. The above inner product \eqref{a4} can therefore be expressed alternatively as

\begin{equation} \label{a6}
\begin{split}
\left( \psi_1 (\hat{x}_1, \hat{x}_2), \psi_2 (\hat{x}_1, \hat{x}_2) \right) & = \int \frac{d^2z}{\pi} \, \bra{z} \psi^\dagger_1 (\hat{x}_1, \hat{x}_2) \psi_2 (\hat{x}_1, \hat{x}_2) \ket{z} \\
 & = \int \frac{d^2x}{2\pi\theta} \, \Tr_{\mathcal{H}_c} \left( \rho_z \psi^\dagger_1 (\hat{x}_1, \hat{x}_2) \psi_2 (\hat{x}_1, \hat{x}_2) \right) ~;~ \rho_z = \dyad{z}{z} \in \mathcal{H}_q \
\end{split}
\end{equation}

Here we have introduced the density matrix $\rho_z \in \mathcal{H}_q$, as viewed from $\mathcal{H}_c$ and can be associated with the pure state $\omega_{\rho_z}$ corresponding to the $\ast$-algebra $\mathcal{H}_q = \mathcal{A}_M$ and defined as a linear functional on $\mathcal{A}_M : \omega_{\rho_z} (a) \in \mathds{C}$ of norm one. As explained in detail in \cite{b7}, here too we shall be working with normal states, so that the states can be represented by density matrices : $\omega_{\rho_z} (a) = \Tr_{\mathcal{H}_c} \left( \rho_z a \right)$. For brevity, therefore, the states will be denoted just by density matrices themselves as $\rho(a) = \Tr \left( \rho a \right)$. Finite distance, \emph{a la} Connes, between the pair of pure states $\rho_0 := \dyad{0}{0}$ and $\rho_z$ has already been computed in \cite{b20} to get

\begin{equation} \label{dis-Moyal}
d\left( \rho_0, \rho_z \right) = \sqrt{2\theta} \abs{z}
\end{equation}

This was also re-derived in a some-what different approach in \cite{b18} by employing the following spectral triple

\begin{equation} \label{moyaltriple}
\mathcal{A}_M = \mathcal{H}_q ~,~ \mathcal{D}_M= \sqrt{\frac{2}{\theta}}\begin{pmatrix}
0&\hat{b}^\dagger\\
\hat{b}&0
\end{pmatrix} ~,~ \mathcal{H}_M = \mathcal{H}_c\otimes \mathds{C}^2 = \text{Span} \bigg\{ \begin{pmatrix}
\ket{\psi}\\
\ket{\phi}\
\end{pmatrix} ~;~ \ket{\psi}, \ket{\phi} \in \mathcal{H}_c \bigg\}
\end{equation}

where the action of the algebra $\mathcal{A}_M$ on the Hilbert space $\mathcal{H}_M$ is given by the diagonal representation $\pi$:

\begin{equation} \label{a9}
\pi(a) = \begin{pmatrix}
a & 0 \\
0 & a \
\end{pmatrix}
\end{equation}

whereas the Dirac operator $\mathcal{D}_M$ acts on $\mathcal{H}_M$ -  the module of spinorial sections from the left. This spectral triple is even as it admits grading or chirality operator $\gamma_M = \sigma_3$ which commutes with $\pi(a)$ for all $a \in \mathcal{A}_M$ and anti-commutes with Dirac operator $\mathcal{D}_M$. The chirality operator splits the Hilbert space into positive and negative sector as $\mathcal{H}_M = \mathcal{H}_+ \oplus \mathcal{H}_- = \left( \mathcal{H}_c \otimes \begin{pmatrix}
1 \\ 0
\end{pmatrix} \right) \oplus \left( \mathcal{H}_c \otimes \begin{pmatrix}
0 \\ 1
\end{pmatrix} \right)$. Let us recall, in this context, that the spectral distance between a pair of states $\rho_1$ and $\rho_2$, which by definition, are a pair of linear functionals of the algebra $\mathcal{A}_M$, is given by

\begin{equation} \label{Connes-dist}
d(\rho_1, \rho_2) = \sup_{a \in B} \abs{\rho_1(a) - \rho_2(a)} ~;~ B = \left\lbrace a \in \mathcal{A}_M  : \|[\mathcal{D}_M, \pi(a)] \|_{op} \leq 1 \right\rbrace
\end{equation}

In \cite{b18}, in particular, the computation of the operator norm was carried out in the eigen-spinor basis $| m \rangle\rangle_\pm \in \mathcal{H}_M$ \eqref{moyaltriple} of the Dirac operator $\mathcal{D}_M$ with eigen-values $\lambda_\pm^{(m)}$, given by

\begin{equation} \label{a11}
\mathcal{D}_M | m \rangle\rangle_\pm = \lambda_m^\pm | m \rangle\rangle_\pm ~;~ | 0 \rangle\rangle_\pm = \begin{pmatrix}
\ket{0} \\
0 \
\end{pmatrix} ~,~ | m \rangle\rangle_\pm = \frac{1}{\sqrt{2}} \begin{pmatrix}
\ket{m} \\
\pm \ket{m-1} \
\end{pmatrix} ~\mathrm{with}~ m \in \{ 1,2,3...\} ~;~ \lambda^{(m)}_\pm = \pm \sqrt{\frac{2m}{\theta}}~\mathrm{with}~ m \in \{ 0,1,2...\}
\end{equation}

They satisfy orthonormality and completeness relation

\begin{equation} \label{a12}
_\pm\langle\langle m | n \rangle\rangle_\pm = \delta_{mn} ~;~ _+\langle\langle m | n \rangle\rangle_- = 0 ~;~ \sum\limits_{m=0}^{\infty} | m \rangle\rangle_\pm ~_\pm \langle\langle m | = \mathds{1}_{\mathcal{H}_M}
\end{equation}

It was also necessary to introduce a projector $\mathds{P}_N$ of rank $(2N+1)$ into $(2N+1)$ dimensional subspace of $\mathcal{H}_M$

\begin{equation} \label{a13}
\mathds{P}_N = \sum\limits_{m=0}^{N} | m \rangle\rangle_\pm ~_\pm \langle\langle m | = \begin{pmatrix}
P_N & 0 \\ 0 & P_{N-1}
\end{pmatrix} ~;~ P_N = \sum\limits_{m=0}^{N}\ket{m}\bra{m}
\end{equation}

with $P_N$ being the projector for $N+1$ dimensional subspace of $\mathcal{H}_c$. Following \cite{b20}, the distance in \eqref{dis-Moyal} was first shown to be the upper bound and subsequently an optimal element

\begin{equation} \label{a14}
a_s = \sqrt{\frac{\theta}{2}} \left( \hat{b} e^{i\alpha} + \hat{b}^\dagger e^{-i\alpha} \right),
\end{equation}

saturating RHS of \eqref{dis-Moyal} as upper bound, so that it can indeed be identified as the true distance. We would like to mention in this context that, although $a_s \notin \mathcal{H}_q = \mathcal{A}_M$, it nevertheless belongs to the multiplier algebra and can be shown \cite{b20} to correspond to the limit point of a sequence, whose elements are in $\mathcal{H}_q$. In an alternative approach, proposed in \cite{b18}, the projected element

\begin{equation} \label{a15}
\mathds{P}_N \pi(a_s) \mathds{P}_N \in \mathcal{H}_q \otimes M_2 (\mathds{C}) ~;~ N \geq 2
\end{equation}

was shown to satisfy the ball condition

\begin{equation} \label{a16}
\| \left[ \mathcal{D}_M, \mathds{P}_N \pi(a_s) \mathds{P}_N \right] \|_{op} = 1~ \forall ~ N \geq 2,
\end{equation}

and shown to yield the correct infinitesimal distance $d(\rho_0, \rho_{dz}) = \sqrt{2\theta} \abs{dz}$ and eventually could be ``integrated" to get \eqref{dis-Moyal} as the finite distance. Although, the conventional notions of points and geodesics do not exist in the Moyal plane, in view of the uncertainty $\Delta x_1 \Delta x_2 \geq \frac{\theta}{2}$, stemming from the noncommutative coordinate sub-algebra \eqref{a1}, the notion of geodesics in the form of a straight line can be retrieved in some sense by explicitly constructing a one-parameter family of pure states $\rho_{zt} := \dyad{zt}{zt}$ with $t \in [0,1]$ interpolating the extremal pure states $\rho_0$ and $\rho_z$, where the triangle in-equality is saturated to an equality : $d(\rho_0, \rho_{zt}) + d(\rho_{zt}, \rho_z) = d(\rho_0, \rho_z)$. This is of course an exception; such a feature does not persist in generic noncommutative spaces. Indeed, for fuzzy sphere $\mathds{S}^2_\ast$ the interpolating states satisfying a similar ``triangle-equality" are necessarily mixed \cite{b18, Liz}. \\
Note that, in order to compute the spectral distance between a pair of pure states on the Moyal plane, we focus our attention to coherent states $|z\rangle$ \eqref{a5} obtained by translating $|0\rangle\in\mathcal{H}_c$ to $|z\rangle = U(z,\bar{z})|0\rangle$ where  $U(z,\bar{z}) := e^{-\bar{z}\hat{b}+z\hat{b}^\dag}$, provides a projective unitary representation of the group of translation. Conversely, by the unitary transformation $U^{-1}(z,\bar{z})$, it is always possible to translate $|z\rangle$ back to $|0\rangle$ such that the density matrix will transform adjointly as

\begin{equation}
\rho=|z\rangle\langle z|\rightarrow U^{-1}(z,\bar{z})\lvert z\rangle\langle z\rvert U(z,\bar{z}) =|0\rangle\langle 0|;
\end{equation}

furnishing a proper representation of the group of translation. This transformation effects the Dirac operator \eqref{moyaltriple} on the single Moyal plane as

\begin{equation} \label{translated_dirac}
\mathcal{D}_M\rightarrow U^{-1}(z,\bar{z})\mathcal{D}_MU(z,\bar{z})=\sqrt{\frac{2}{\theta}}\begin{pmatrix}
0&\hat{b}^\dagger-\bar{z}\\
\hat{b}-z&0
\end{pmatrix}
\end{equation}

where  $(\hat{b}-z)|z\rangle =0$  can be rewritten as 

\begin{equation} \label{translation}
\hat{\tilde{b}}|\tilde{0}\rangle = 0; ~\hat{\tilde{b}}=\hat{b}-z,~ |\tilde{0}\rangle=|z\rangle
\end{equation}

Similarly, we have $(\hat{b}^\dagger-\bar{z})|z\rangle = \hat{\tilde{b}}^\dagger|\tilde{0}\rangle= |\tilde{1}\rangle$,  
and higher states can be constructed by repeated actions of the new creation operators $\hat{\tilde{b}}^\dagger$:

\begin{equation}
\ket{\tilde{n}} =  \frac{1}{\sqrt{n!}}(\hat{\tilde{b}}^\dagger )^n \ket{\tilde{0}},
\end{equation}

 which shows that the new Fock space can be defined by translating the ``vacuum" by a c-number, which would then be annihilated by the ``translated" lowering operator. Finally note that the change in $\mathcal{D}_M$ under translation \eqref{translated_dirac} is again given by an element $-\sqrt{\frac{2}{\theta}}\begin{pmatrix}
 0 & \bar{z} \\ z & 0
 \end{pmatrix} \in M_2(\mathds{C}) $, so that the ball condition $B$ occurring in \eqref{Connes-dist} and hence the spectral distance remains invariant under translation \cite{b18}.

\section{Doubling spectral triple}  \label{sec3}

\begin{wrapfigure}{l}{8cm}
\includegraphics[width=8cm]{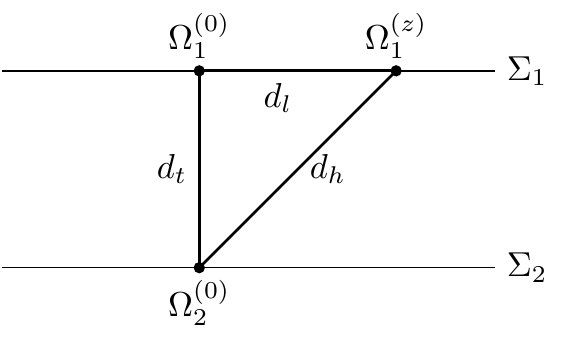}
\caption{$\mathds{R}_{*}^2 \cup \mathds{R}_{*}^2$, Space associated with double Moyal plane.}
\end{wrapfigure}

In non-commutative geometry, the notion of usual manifold is generalized by the spectral triple $T=(\mathcal{A},\mathcal{H},\mathcal{D})$ which obey some axioms \cite{Connes1} and if we have two spectral triples $T_1$ and $T_2$ associated with two spaces, then the composite spectral triple denoted also by the tensor product notation $T_1\otimes T_2$ \footnote{Although $T_1$ and $T_2$ are not vector spaces themselves, we still use this convention of notation as in  \cite{Martinetti3}} generalizes the notion of fibre bundle over a manifold \cite{Martinetti3}. If one of the spectral triples is taken to be that of finite discrete space, say, for example the simplest known finite discrete space of the commutative two-point space for which the spectral triple is given by (A brief review of this two-point space has been provided in the appendix, where the relevant notations are also introduced):

\begin{equation}
T_2 = \left( \mathcal{A}_2=\mathds{C}^2 \simeq M^d_2(\mathds{C}) , ~~\mathcal{H}_2=\mathds{C}^2, ~~ \mathcal{D}_2=\begin{pmatrix}
0 & \Lambda \\ \bar{\Lambda} & 0
\end{pmatrix} \right) ~~;~~ \Lambda \in \mathds{C}, \label{2pt-triple}
\end{equation}

then, one can take the product of any even spectral triple $T_1$ with that of two-point space so that the resulting spectral triple can be written as,

\begin{equation} \label{prod_triple}
T := T_1\otimes T_2= (\mathcal{A}=\mathcal{A}_1\otimes \mathds{C}^2,\mathcal{H}=\mathcal{H}_1\otimes \mathds{C}^2,\mathcal{D}=\mathcal{D}_1\otimes \mathds{1}_2+\gamma_1\otimes\mathcal{D}_2 ) ~;~ \gamma_1 ~\mathrm{is~the~grading~operator~of~T_1}
\end{equation}

describes the space $M\cup M$ where $M$ is the space associated with the spectral triple $T_1$. In particular, if the spectral triple $T_1$ is that of Moyal plane \eqref{moyaltriple} i.e. $M = \mathds{R}_{*}^2$, with grading operator $\gamma_1 = \gamma_M = \sigma_3$, then the product space $T := T_1\otimes T_2$, is known as double Moyal plane the spectral triple of which is given by

\begin{equation} \label{spec_trip}
\mathcal{A}_T = \mathcal{H}_q\otimes M^d_2(\mathds{C}),~ \mathcal{H}_T = (\mathcal{H}_c\otimes\mathds{C}^2)\otimes\mathds{C}^2, ~ \mathcal{D}_T = \mathcal{D}_M\otimes\mathds{1}_2+\sigma_3 \otimes\mathcal{D}_2.
\end{equation}

Here the subscript $T$ stands for total. Further the total grading operator here takes the form 

\begin{equation} \label{gamma_total}
\gamma_T = \gamma_M \otimes \gamma_2 = \sigma_3 \otimes \sigma_3,
\end{equation}

as the grading operator for the two-point space is same as that for the Moyal plane i.e. $\gamma_2= \gamma_M = \sigma_3$. The pure states in double Moyal plane (see Figure 1), between which we shall be computing distances, also comes in tensor product form

\begin{equation}
\Omega_i^{(z)} = \rho_z\otimes\omega_i, ~~~ i \in \{1,2\}, \label{unionstate}
\end{equation} 

where $\omega_1$ and $\omega_2$ are the only two pure states of two point space (see \eqref{2-pt-states} in Appendix). The fact that the ``composite" pure states $\Omega_i^{(z)}$ also persists to be pure is because of the fact that the algbera $\mathcal{A}_2 = \mathds{C}^2$ is abelian. In Figure 1, the two copies of Moyal plane $\left(\mathds{R}_{*}^2\right)$ are denoted by $\Sigma_i$ based on the pure states $\omega_i$. It was shown in \cite{b20}, for unital double Moyal plane, that the `transverse' distance $d_t$ between a state $\rho_z$ of the single Moyal plane and its ``clone" belonging to the other Moyal plane to be same as that of the distance between states $\omega_1$ and $\omega_2$ of two-point space (see appendix)\footnote{This distance was used in \cite{b20} to reconcile the non-vanishing nature of ``quantum length" between a state and itself \cite{B} with spectral distance between a state and it's clone, thereby identifying $\frac{1}{\abs{\Lambda}} \sim \sqrt{\theta} \sim L_p$ -Plank length. This issue, however, is beyond the scope of this paper.}

\begin{equation}
d_t\left(\Omega_1^{(z)},\Omega_2^{(z)} \right) = d_t \left( \rho_z\otimes\omega_1, \rho_z\otimes\omega_2 \right) = d_{\mathcal{D}_2}\left(\omega_1, \omega_2 \right) = \frac{1}{|\Lambda|}. \label{trans_dis}
\end{equation}

Analogously, the longitudinal distance $d_l$, computed between states $\rho_0$ and $\rho_z$ belonging to the same copy of Moyal plane $\Sigma_i$ amounted to computing the distance between $\rho_0$ and $\rho_z$ on the same Moyal plane itself

\begin{equation} \label{long_dis}
 d_l\Big(\Omega_i^{(0)},\Omega_i^{(z)} \Big) = d_{\mathcal{D}_M}(\rho_0,\rho_z) = \sqrt{2\theta}\abs{z}.
\end{equation}

Distance, $d_h$, between states like $\Omega_1^{(z)}$ and $\Omega_2^{(0)}$, as shown in Figure 1, are called `hypotenuse' distance. In case of commutative and unital spectral triple, these distances obey the Pythagoras equality

\begin{equation} \label{Pytha}
\Big\{d_h(\Omega_1=\rho\otimes\omega_1,\Omega_2=\rho'\otimes\omega_2)\Big\}^2=\Big\{d_t(\Omega_1=\rho\otimes\omega_1,\Omega_2=\rho\otimes\omega_2) \Big\}^2+\Big\{d_l(\Omega_1=\rho\otimes\omega_1,\Omega_2=\rho'\otimes\omega_1) \Big\}^2.
\end{equation}

This Pythagoras theorem is shown to obey even for the non-commutative but unital spectral triple \cite{Martinetti3} and by unitizing the algebra associated with the Moyal plane, it was shown that Pythagoras theorem is obeyed on the double unitized Moyal plane \cite{b20}. Here, one of our aims is to compute the above spectral distances on the double Moyal plane, in the  context of Hilbert-Schmidt operatorial formulation of non-commutative quantum mechanics \cite{b2}, by making use of Dirac eigen-spinors and verify Pythagoras equality without unitizing the algebra $\mathcal{H}_q$.
\\

Just as the algebra elements of Moyal plane acts on it's Hilbert space elements through diagonal representation \eqref{a9}, a generic element of $a_T = |\psi)\otimes\begin{pmatrix}
\lambda_1&0\\
0&\lambda_2
\end{pmatrix}\in\mathcal{A}_T = \mathcal{H}_q\otimes M_2^d(\mathds{C})$, where $ M_2^d(\mathds{C})$ is the c-valued $2\times 2$ diagonal matrix (see (\ref{a-in-C2}) in appendix) acts on a generic element $|\Phi\rangle_T = \begin{pmatrix}
|\phi_1\rangle\\
|\phi_2\rangle
\end{pmatrix}\otimes\begin{pmatrix}
\mu_1\\
\mu_2
\end{pmatrix}$  $\in\mathcal{H}_T =(\mathcal{H}_c\otimes\mathds{C}^2)\otimes\mathds{C}^2$ (with  $\lambda_1,\lambda_2,\mu_1,\mu_2 \in\mathds{C}$) as
\begin{equation}
\pi(a_T)|\Phi\rangle_T = \bigg\{\begin{pmatrix} |\psi)&0\\0&|\psi)\end{pmatrix}\otimes\begin{pmatrix}\lambda_1&0\\ 0&\lambda_2\end{pmatrix}\bigg\}\bigg\{\begin{pmatrix}
|\phi_1\rangle\\
|\phi_2\rangle
\end{pmatrix}\otimes\begin{pmatrix}
\mu_1\\
\mu_2
\end{pmatrix}\bigg\} = \begin{pmatrix}
|\psi)|\phi_1\rangle\\|\psi)|\phi_2\rangle
\end{pmatrix}\otimes\begin{pmatrix}
\lambda_1\mu_1\\\lambda_2\mu_2
\end{pmatrix}.\label{piat}
\end{equation}

Moreover, the total Dirac operator $\mathcal{D}_T$ \eqref{spec_trip} is obtained by substituting the forms of $\mathcal{D}_M$ from \eqref{moyaltriple} and $\mathcal{D}_2$ \eqref{2-point-triple} in the appendix to get,

\begin{eqnarray} \label{dirac_op}
\mathcal{D}_T &=&\sqrt{\frac{2}{\theta}}\begin{pmatrix}
0&\hat{b}^\dagger\\
\hat{b}&0
\end{pmatrix}\otimes\begin{pmatrix}
1&0\\
0&1
\end{pmatrix} +\begin{pmatrix}
1&0\\
0&-1
\end{pmatrix}\otimes \begin{pmatrix}
0 & \Lambda\\
\bar{\Lambda} & 0
\end{pmatrix}.
\end{eqnarray}

By considering the phase of $\Lambda$ to be $\phi$ i.e. $\Lambda = \abs{\Lambda}e^{i\phi}$, we can make a unitary transformation of above Dirac operator as $\mathcal{D}_T \rightarrow U(\phi) \mathcal{D}_T U^\dagger (\phi)$, where $U(\phi)$ is given by

\begin{equation} \label{unitary_op}
U (\phi) = \begin{pmatrix}
\mathds{1}_{\mathcal{H}_q} & 0 \\ 0 & \mathds{1}_{\mathcal{H}_q}
\end{pmatrix} \otimes \begin{pmatrix}
e^{\frac{-i\phi}{2}} & 0 \\ 0 & e^{\frac{i\phi}{2}}
\end{pmatrix},
\end{equation}

so that the Dirac operator becomes

\begin{eqnarray}
\mathcal{D}_T &=&\sqrt{\frac{2}{\theta}}\begin{pmatrix}
0&\hat{b}^\dagger\\
\hat{b}&0
\end{pmatrix}\otimes\begin{pmatrix}
1&0\\
0&1
\end{pmatrix} + \abs{\Lambda} \begin{pmatrix}
1&0\\
0&-1
\end{pmatrix}\otimes \begin{pmatrix}
0 & 1\\
1 & 0
\end{pmatrix} \label{totalDiracop}
\end{eqnarray}

The first slot of $U(\phi)$ in \eqref{unitary_op}, which is the representation of $\mathds{1}_{\mathcal{H}_q}$ is, however, a bit troubling since the Moyal plane algebra $\mathcal{A}_M = \mathcal{H}_q$ is not-unital : $\mathds{1}_{\mathcal{H}_q} \not\in \mathcal{H}_q$. The way out is provided by the sequence of projection operators $\mathds{P}_N$ \eqref{a13}, all of which lies in $\mathcal{B}(\mathcal{H}_M)$ \eqref{moyaltriple} and can be taken to reproduce the identity element in the limit i.e. 

\begin{equation} \label{limit}
\lim_{N \rightarrow \infty} \mathds{P}_N =\pi( \mathds{1}_{\mathcal{H}_q}),
\end{equation}

in the sense that the composite operators formed by multiplying any compact operators, like a Hilbert-Schmidt operator $\in \mathcal{H}_q$ by $\mathds{P}_N$ is same as that of multiplication with identity operator in the limit $N \rightarrow \infty$.  Note that, in order to arrive at a ``unitary equivalent" spectral triple i.e. $(\mathcal{A}_T, \mathcal{H}_T, U(\phi) \mathcal{D}_T U^\dagger (\phi)$), with $U(\phi)$ as in \eqref{unitary_op}, one needs to make a simultaneous unitary transformation of $\pi(a_T)$ using this $U(\phi)$ at the same time. This simultaneous transformation ensures that the ball condition ($B$) in \eqref{Connes-dist} remains invariant thus yielding the same distance. It is, however, quite straightforward to verify that $\pi(a_T)$ as given in \eqref{piat} remains invariant under such unitary transformation and therefore we need not worry about this issue any more. From this point onward, unless mentioned otherwise we will therefore take $\Lambda \in \mathds{R}$ and positive $\Lambda > 0$, without loss of generality in order to avoid the cluttering of notations. Moreover, such unitary equivalent spectral triple are a part of several other transformation which preserves the metric properties of the triple (see \cite{Martinetti3}). Before concluding this section, we would like to mention that it is also possible to retrieve individual spectral triples $T_1$ or $T_2$ along with their respective distances from the composite one $T$ \eqref{prod_triple}, provided certain conditions are satisfied. For the spectral triple \eqref{spec_trip} this conditions are indeed satisfied, as we shall exhibit now. To that end, we begin with a brief review of the concept of restricted spectral triple as introduced in \cite{Martinetti3}. For a given spectral triple $(\mathcal{A}, \mathcal{H}, \mathcal{D})$, let us consider the action of a self-adjoint element $\rho \in \mathcal{A}$, satisfying the property of projector: $\rho^2 = \rho = \rho^*$, on an arbitrary algebra elements $a \in \mathcal{A}$ through following map

\begin{equation}
\alpha_{\rho} : \mathcal{A} \rightarrow \mathcal{A};~ a \mapsto \alpha_{\rho}(a) = \rho a \rho.
\end{equation}

This transformation gives rise to following ``restricted" spectral triple

\begin{equation} \label{restricted_triple}
\mathcal{A}^{(\rho)} = \alpha_{\rho} (\mathcal{A}), ~\mathcal{H}^{(\rho)} = \pi(\rho)\mathcal{H},~  \mathcal{D}^{(\rho)} = \pi(\rho)\mathcal{D} \pi(\rho), 
\end{equation} 

with the representation $\pi$ being restricted to $\pi|_{\mathcal{H}^{(\rho)}}$. For a pair of pure states $\omega_1$ and $\omega_2$ of $\mathcal{A}^{(\rho)}$ the corresponding spectral distance then remains unaffected by projection (see e.g. Lemma 1 in \cite{Martinetti3}) i.e.

\begin{equation} \label{inv_dist}
d^{(\rho)} (\omega_1, \omega_2) = d(\omega_1 \circ \alpha_\rho, \omega_2 \circ \alpha_\rho)~\forall~ \omega_1, \omega_2 \in \mathcal{P}(\mathcal{A}^{(\rho)}),
\end{equation}

provided 

\begin{equation} \label{proj_cond}
[\mathcal{D}, \pi(\rho)] = 0.
\end{equation}

This condition then implies that $\pi(\rho)$ should then correspond to the projection operators and should be built out of the eigenspinors of the Dirac operator. In the case of the spectral triple \eqref{spec_trip}, for example, this should involve $\mathds{P}_N$ \eqref{a13} and $\omega_i$ \eqref{2-pt-states}. Indeed, it can be verified in a straightforward manner that the following projection operators $P^{(trans)}_{T(0)}$ and $P^{(long)}_{T(i)}(N)$, defined as 

\begin{align}\label{proj_new}
\begin{split}
P^{(trans)}_{T(0)} := \mathds{P}_0 \otimes \mathds{1}_2 = \begin{pmatrix}
|0\rangle \langle 0| & 0 \\ 0 & 0
\end{pmatrix} \otimes \mathds{1}_2 \in \mathcal{A}_T  \\ 
P^{(long)}_{T(i)}(N) := \mathds{P}_N \otimes \omega_i = \begin{pmatrix}
P_N & 0 \\ 0 & P_{N-1}
\end{pmatrix} \otimes \omega_i ~;~ i = 1,2  
\end{split}
\end{align}

can be used to construct the following two spectral triples from \eqref{spec_trip}. The first one yields

\begin{equation} \label{proj_triple} 
P^{(trans)}_{T(0)} \mathcal{A}_T P^{(trans)}_{T(0)} = \begin{pmatrix}
|0\rangle \langle 0| & 0 \\ 0 & 0
\end{pmatrix} \otimes M^d_2 (\mathds{C})~;~ P^{(trans)}_{T(0)} \mathcal{H}_T = \begin{pmatrix}
|0\rangle \\ 0
\end{pmatrix} \otimes \mathds{C}^2~;~ P^{(trans)}_{T(0)} \mathcal{D}_T P^{(trans)}_{T(0)} = \mathds{P}_0 \otimes \mathcal{D}_2
\end{equation}

which is clearly identical to the one involving two-point space \eqref{2pt-triple}. Note that in this case the role of the representation $\pi$ becomes redundant. Essentially following the same approach one can also retrieve the unitized version of the spectral triple for single Moyal plane \eqref{moyaltriple} by making use of the projector $P^{(long)}_{T(i)}(N)$ in the limit $N \rightarrow \infty$. Note that for any finite $N~(0<N<\infty)$ this does not belong to $\mathcal{A}_T$ and this stems from the fact that, the projector $\mathds{P}_N$ \eqref{a13} by itself, can not be identified with the diagonal representation $\pi$ of $\mathcal{A_M} = \mathcal{H}_q$ \eqref{moyaltriple}. It is now quite clear that either of the projectors \eqref{proj_new} commutes with the total Dirac operator $\mathcal{D}_T$ \eqref{spec_trip}:

\begin{equation}
[\mathcal{D}_T,  P^{(trans)}_{T(0)} ] = [\mathcal{D}_T, P^{(long)}_{T(i)}(N)] = 0,
\end{equation}

which is precisely the condition \eqref{proj_cond} so that the results (\ref{trans_dis}, \ref{long_dis}), regarding transverse and longitudinal distances follow trivially from \eqref{inv_dist}. Further, the fact that the RHS of \eqref{trans_dis} is independent of `z' can be seen easily by considering a slight variant of the projector $P^{(trans)}_{T(0)}$ viz. 

\begin{equation} \label{proj_translated}
P^{(trans)}_{T(z)} = \mathds{P}_{\tilde{0}} \otimes \mathds{1}_2 = \begin{pmatrix}
| \tilde{0} \rangle \langle \tilde{0} | & 0 \\ 0 & 0
\end{pmatrix} \otimes \mathds{1}_2
\end{equation} 

where the states $| \tilde{0} \rangle \langle \tilde{0}|$ can be thought to be anchored to the ``shifted" origin (in the spirit of Gelfand and Naimark) as discussed earlier \eqref{translation} and finally invoking the translational invariance of spectral distance of coherent states, parametrized by the complex plane (see below \eqref{translation}).

\section{Construction of eigen spinors of the total Dirac operator $\mathcal{D}_T$} \label{sec4}

Looking at the total Dirac operator \eqref{totalDiracop} we realize that it's eigenspinors would belong to a space spanned by the tensor product of Moyal plane eigenspinors \eqref{a11} on the left slot and eigenspinors $\frac{1}{\sqrt{2}} \begin{pmatrix}
1\\ \pm 1 \
\end{pmatrix}$ of $\sigma_1 = \begin{pmatrix}
0 & 1\\ 1 & 0 \
\end{pmatrix}$ on the right slot. One is immediately tempted to work with irreducible subspaces of eigenspinors viz. spin up ($\uparrow$) subspace Span$\big\{ V^{(m)}_{++}, V^{(m)}_{-+} \big\}$ formed by tensoring with $\ket{\uparrow} := \frac{1}{\sqrt{2}} \begin{pmatrix}
1\\ 1 \
\end{pmatrix}$ with $\sigma_1 \ket{\uparrow} = + \ket{\uparrow}$ and spin down ($\downarrow$) subspace Span$\big\{ V^{(m)}_{+-}, V^{(m)}_{--} \big\}$ formed by tensoring with $\ket{\downarrow} := \frac{1}{\sqrt{2}} \begin{pmatrix}
1\\ -1 \
\end{pmatrix}$ with $\sigma_1 \ket{\downarrow} = - \ket{\downarrow}$ where the (unnormalised) spinors $V^{(m)}_{\pm\pm}$ are given by

\begin{equation} \label{basis}
V^{(m)}_{\pm\pm} = \begin{pmatrix}
\ket m \\ \pm\ket {m-1}
\end{pmatrix}\otimes \begin{pmatrix}
1 \\ \pm 1
\end{pmatrix}.
\end{equation}

Now considering an arbitrary linear combination of spinors of spin up subspace i.e. $C_m^1 V^{(m)}_{++} + C_m^2 V^{(m)}_{-+}$ and demanding that it should be an eigenspinor of total Dirac operator \eqref{totalDiracop} we obtain following eigenvalue equation

\begin{equation} \label{mat1}
\begin{pmatrix}
\sqrt{\frac{2m}{\theta}} & \Lambda \\
\Lambda & -\sqrt{\frac{2m}{\theta}} 
\end{pmatrix}\begin{pmatrix}
C_m^1 \\
C_m^2 
\end{pmatrix} = \lambda^{(m)} \begin{pmatrix}
C_m^1 \\
C_m^2 
\end{pmatrix},
\end{equation}

whose eigenvalues are found to be

\begin{equation} \label{eigenvalue}
\lambda^{(m)}_{\pm} = \pm\Lambda\sqrt{\kappa m +1}~;~ \kappa = \frac{2}{\theta\Lambda^2},
\end{equation} 

and the eigenspinors are (upto appropriate normalization)

\begin{equation} \label{eig-spinor1}
\ket{\Phi^{(m)}_{+\uparrow}} = V^{(m)}_{-+} + \left(\sqrt{\kappa m + 1} + \sqrt{\kappa m}\right)V^{(m)}_{++} ~~;~~ \ket{\Phi^{(m)}_{-\uparrow}} = V^{(m)}_{-+} - \left(\sqrt{\kappa m + 1} - \sqrt{\kappa m}\right)V^{(m)}_{++}.
\end{equation}

Here the subscript $+$ and $-$ refers to the eigen value being $\lambda^{(m)}_+$ and $\lambda^{(m)}_-$ respectively. Repeating the same process for spin down subspace we obtain the same set of eigenvalues \eqref{eigenvalue} and following eigenspinors

\begin{equation} \label{eig-spinor2}
\ket{\Phi^{(m)}_{+\downarrow}} = V^{(m)}_{--} - \left(\sqrt{\kappa m + 1} + \sqrt{\kappa m}\right)V^{(m)}_{+-}~~;~~ \ket{\Phi^{(m)}_{-\downarrow}} = V^{(m)}_{--} + \left(\sqrt{\kappa m + 1} - \sqrt{\kappa m}\right)V^{(m)}_{+-}.
\end{equation}

However, these subspaces are not invariant under the action of algebra $\mathcal{A}_T$ \eqref{spec_trip} as some of the algebra elements (e.g. in case of transverse distance) mixes the eigenspinors of two subspaces and one has to work with all the four of them anyhow. It turns out that working with following (symmetric) linear combination of eigen spinors, rather than the ones obtained earlier (\ref{eig-spinor1}-\ref{eig-spinor2}) simplifies our calculation drastically.

\begin{align} \label{ei-spi}
\begin{split}
\ket{\Psi^{(m)}_+} = \left(\sqrt{\kappa m + 1} - \sqrt{\kappa m}\right) \ket{\Phi^{(m)}_{+\uparrow}} + \ket{\Phi^{(m)}_{+\downarrow}} ~;~ \ket{\Psi^{(m)}_-} = \ket{\Phi^{(m)}_{-\downarrow}} - \left(\sqrt{\kappa m} + \sqrt{\kappa m + 1}\right) \ket{\Phi^{(m)}_{-\uparrow}} \\
\ket{\widetilde{\Psi}^{(m)}_+} = \ket{\Phi^{(m)}_{+\uparrow}} - \left(\sqrt{\kappa m + 1} - \sqrt{\kappa m}\right) \ket{\Phi^{(m)}_{+\downarrow}} ~;~ \ket{\widetilde{\Psi}^{(m)}_-} = \ket{\Phi^{(m)}_{-\uparrow}} + \left(\sqrt{\kappa m} + \sqrt{\kappa m + 1}\right) \ket{\Phi^{(m)}_{-\downarrow}}
\end{split}
\end{align}

One can write these eigen spinors, after normalization, in the following compact form.

\begin{align} \label{eig-spi-1}
\begin{split}
\ket{\Psi^{(m)}_{\pm}} = N_m \left[ V^{(m)}_{++} + V^{(m)}_{--} \pm V^{(m)}_{-+} \left(\sqrt{\kappa m + 1} \mp \sqrt{\kappa m}\right) \mp V^{(m)}_{+-}\left(\sqrt{\kappa m + 1} \pm \sqrt{\kappa m}\right) \right] \\
\ket{\widetilde{\Psi}^{(m)}_{\pm}} = N_m \left[ V^{(m)}_{+-} + V^{(m)}_{-+} \pm V^{(m)}_{++}\left(\sqrt{\kappa m + 1} \pm \sqrt{\kappa m}\right) \mp V^{(m)}_{--}\left(\sqrt{\kappa m + 1} \mp \sqrt{\kappa m}\right)\right], ~~ N_m = \frac{1}{4\sqrt{\kappa m + 1}}
\end{split}
\end{align}

where $m \in \{ 1,2,3... \}$ and $N_m$ is the normalization constant, which takes this simple form because of the symmetry of the new eigenspinors. Moreover the case of $m=0$ is 2 dimensional, unlike the eigenspinors in \eqref{eig-spi-1} which are 4 dimensional for each $m$, and can be written after normalization as

\begin{equation} \label{eig-spi-2}
\ket{\Psi^{(0)}_{\pm}} = \frac{1}{\sqrt{2}}\begin{pmatrix}
|0\rangle \\ 0
\end{pmatrix}\ \otimes \begin{pmatrix}
1 \\ \pm 1
\end{pmatrix}
\end{equation}

Here again the subscript $\pm$ represent the eigenvalues $\lambda^{(0)}_{\pm} =\pm\Lambda$. The whole set (\ref{eig-spi-1}, \ref{eig-spi-2}) furnishes an orthonormal basis for the entire space:

\begin{align} 
\begin{split}
\Big\langle \Psi^{(m)}_{\pm} \Big| \Psi^{(n)}_{\pm} \Big\rangle = \Big\langle \widetilde{\Psi}^{(m)}_{\pm} \Big| \widetilde{\Psi}^{(n)}_{\pm} \Big\rangle = \delta^{mn};~ \Big\langle \Psi^{(0)}_{\pm} \Big| \Psi^{(0)}_{\pm} \Big\rangle = 1; ~ m,n \in \{ 1,2,3... \} \\
\Big\langle \Psi^{(m)}_{\pm} \Big| \Psi^{(n)}_{\mp} \Big\rangle = \Big\langle \widetilde{\Psi}^{(m)}_{\pm} \Big| \widetilde{\Psi}^{(n)}_{\mp} \Big\rangle = \Big\langle \widetilde{\Psi}^{(m)}_{\pm} \Big| \Psi^{(n)}_{\pm} \Big\rangle = \Big\langle \widetilde{\Psi}^{(m)}_{\pm} \Big| \Psi^{(n)}_{\mp} \Big\rangle = 0  \\
\Big\langle \Psi^{(m)}_{\pm} \Big| \Psi^{(0)}_{\pm} \Big\rangle = \Big\langle \Psi^{(m)}_{\pm} \Big| \Psi^{(0)}_{\mp} \Big\rangle = \Big\langle \widetilde{\Psi}^{(m)}_{\pm} \Big| \Psi^{(0)}_{\pm} \Big\rangle = \Big\langle \widetilde{\Psi}^{(m)}_{\pm} \Big| \Psi^{(0)}_{\mp} \Big\rangle = 0.
\end{split}
\end{align}

\section{Computing distances in double Moyal plane} \label{section5}

We will compute, in this section, exact distances for all three cases and reproduce the results obtained by Martinetti et al. \cite{b20} including the Pythagoras equality for the double Moyal plane, by making use of Dirac eigenspinors (\ref{eig-spi-1}-\ref{eig-spi-2}) that we have just constructed. This will involve making right ansatz regarding the structure of optimal algebra elements belonging to $\mathcal{A}_T$ \eqref{spec_trip}. The necessary hint can be obtained from the corresponding structure of \eqref{a14} for single Moyal plane. In an alternative approach, we show how the projection operators built out of these same Dirac eigenspinors can be made use of, instead of unitizing the algebra $\mathcal{A}$ as in \cite{b20}, to compute transverse distance. 

\subsection{Longitudinal distance} \label{sec5}

To start with, we consider a generic algebra element $a_T \in \mathcal{A}_T = \mathcal{H}_q\otimes M_2^d(\mathds{C})$ which can be written as $a_T = a \otimes a_2$, where $a \in \mathcal{H}_q$ and $a_2 := \begin{pmatrix}
c_1 & 0 \\ 0 & c_2
\end{pmatrix} \in \mathds{C}^2$. Note that for optimal algebra element to be self-adjoint $a_T^\dagger = a_T$ \cite{b20}, we set $a^\dagger = a$ and restrict $c_1, c_2 \in \mathds{R}$. To compute the longitudinal distance $d_{PQ} = d_{P'Q'}$ as shown in Figure 2, we substitute the states $\Omega^{(z)}_i$ and $\Omega^{(0)}_i$ from \eqref{unionstate} in the Connes distance formula \eqref{Connes-dist}, with algebra element $a_T$, to obtain

\begin{eqnarray} \label{exact_dis-1}
d_l\left( \Omega^{(z)}_i, \Omega^{(0)}_i \right) &=& \sup_{a_T \in B_T} \abs{\Omega^{(z)}_i(a_T) - \Omega^{(0)}_i(a_T)} \\
&=& \sup_{a_T \in B_T} \abs{ \Tr_M \left(d\Omega_i a_T\right)};~d\Omega_i = \Omega^{(z)}_i - \Omega^{(0)}_i  \label{dis-1.2} \\
&=& \sup_{a_T \in B_T} \abs{c_i}\cdot \abs{ \Tr_c \left((\rho_z - \rho_0) a\right)}    \\
&=& \sup_{a_T \in B_T} \abs{c_i}\cdot \abs{ \rho_{z}(a) - \rho_0(a)},  \label{dis-1.1}
\end{eqnarray}

Also the ball condition $B_T$ is built from total Dirac operator $D_T$ \eqref{totalDiracop} i.e. $\|[\mathcal{D}_T,\pi(a_T)]\|_{op}\leq 1$ of double Moyal plane and upon simplification looks like

\begin{equation} \label{ball_new}
[\mathcal{D}_T,\pi(a_T)] = [\mathcal{D}_M,\pi(a)]\otimes a_2 + a(c_1 - c_2)\sigma_3 \otimes \mathcal{D}_2
\end{equation} 

It is clear from \eqref{dis-1.1}, by the symmetry requirement that the longitudinal distance has to be same on the two different sheets of double Moyal plane, that $\abs{c_1} = \abs{c_2}$ or $c_1 = \pm c_2$. Note that the condition $c_1 = c_2 = X$ (say) renders the total ball condition $B_T$ to be same as the ball condition $B_M$ of single Moyal plane alone as is clear from \eqref{ball_new}, and we recover the Moyal plane distance \eqref{dis-Moyal}. The case of $c_1 = -c_2$ gives distance which is lower than $\sqrt{2\theta}\abs{z}$ \eqref{dis-Moyal}, as we will demonstrate shortly using an algebra element $a_l$, and therefore will be rejected. \\

\begin{wrapfigure}{l}{8cm}
\includegraphics[width=8cm]{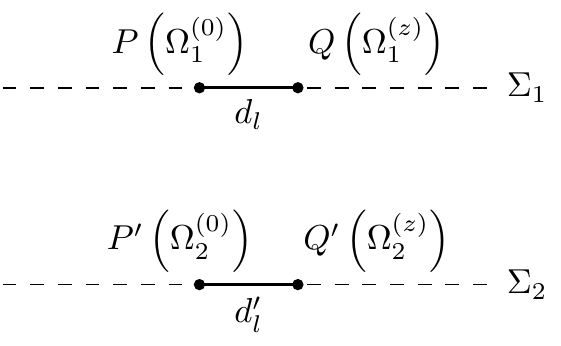}
\caption{Different states belonging to same Moyal plane.}
\end{wrapfigure}

Let us now take as an ansatz the optimal element
\begin{equation} \label{A1}
a_s^{(l)} := \left( b + b^\dagger \right) \otimes a_2 \in \mathcal{A}_T
\end{equation}
(superscript `l' stands for longitudinal)for the computation of longitudinal distances. This is motivated from \eqref{a14} with the phase $\alpha$ coming from $z = \abs{z}e^{i\alpha}$ set to $\alpha = 0$ by invoking the rotational symmetry on Moyal plane, so that only the distance along the real axis is computed. Then a simple calculation yields following trace

\begin{equation} \label{tr_1}
\Tr_M\left( d\Omega_i \left( a_s^{(l)} \right) \right) = 2c_iz.
\end{equation}

Now the matrix representation of $\left[\mathcal{D}_T,\pi\left( a_s^{(l)} \right) \right]$ would be infinite dimensional and one can project it on some finite dimensional subspace (as for the single Moyal plane in \cite{b18}) in order to get hold of the ball condition $B_T$. There is a natural way to acheive this by using projection operators $\mathcal{P}_N \in \left( \mathcal{H}_q\otimes M_2(\mathds{C}) \right)\otimes M_2(\mathds{C}) $, constructed from the Dirac eigenspinors (\ref{eig-spi-1},\ref{eig-spi-2}) in the following way

\begin{eqnarray} \label{proj-2}
\mathcal{P}_N = \ket {\Psi^{(0)}_+}\bra {\Psi^{(0)}_+} + \ket {\Psi^{(0)}_-}\bra {\Psi^{(0)}_-} + \sum^{N}_{m=1} & \bigg[& \ket {\Psi^{(m)}_+}\bra {\Psi^{(m)}_+} + \ket {\Psi^{(m)}_-}\bra {\Psi^{(m)}_-}  \\
&~&  \ket{\widetilde{\Psi}^{(m)}_+}\bra {\widetilde{\Psi}^{(m)}_+} + \ket{\widetilde{\Psi}^{(m)}_-}\bra {\widetilde{\Psi}^{(m)}_-} \bigg]  \nonumber
\end{eqnarray}

These projection operators $\mathcal{P}_N$ belong to the same space where $\pi(a_T)~\forall~ a_T \in \mathcal{A}_T$ belongs to and which is $\left( \mathcal{H}_q\otimes M_2^d(\mathds{C}) \right)\otimes M_2^d(\mathds{C})$ and after explicit calculation, quite remarkably, it splits as 

\begin{equation} \label{proj-3}
\mathcal{P}_N = \mathds{P}_N\otimes \mathds{1}_2,
\end{equation}
 
where the projection operator $\mathds{P}_N$ is given by \eqref{a13}. We can thus compute various finite dimentional (projected) matrix representation of $\left[\mathcal{D}_T,\pi\left( a_s^{(l)} \right)\right]$ using \eqref{proj-2} in the form of $\left[\mathcal{D}_T,\mathcal{P}_N\pi\left( a_s^{(l)} \right) \mathcal{P}_N\right]$. For $\mathcal{P}_2$ and $c_1 = c_2 = X$ (one of the two cases of $c_1 = \pm c_2$ as discussed earlier), the matrix $\left[\mathcal{D}_T,\mathcal{P}_2\pi\left( a_s^{(l)} \right)\mathcal{P}_2\right] =: M_l$ (say) takes following form

\begin{eqnarray} \label{mat2}
\footnotesize
\arraycolsep=0.5pt 
\medmuskip = 2mu
M_l \thickmuskip = 0.1mu = \begin{pmatrix}
0 & 0 & \gamma_-\delta_3 & \gamma_+\delta_4 & \gamma_-\delta_1 & \gamma_+\delta_2 & 0 & 0 & 0 & 0 \\
0 & 0 & -\gamma_+\delta_4 & -\gamma_-\delta_3 & -\gamma_+\delta_2 & -\gamma_-\delta_1 & 0 & 0 & 0 & 0 \\
-\gamma_-\delta_3 & \gamma_+\delta_4 & 0 & 0 & 0 & 0 & \eta(\beta_- - \epsilon_+) & -\eta(\beta_+ + \epsilon_-) & -\eta(\beta_- + \epsilon_-) & \eta(\beta_+ - \epsilon_+) \\
-\gamma_+\delta_4 & \gamma_-\delta_3 & 0 & 0 & 0 & 0 & \eta(\beta_+ + \epsilon_-) & -\eta(\beta_- - \epsilon_+) & -\eta(\beta_+ - \epsilon_+) & \eta(\beta_- + \epsilon_-) \\
-\gamma_-\delta_1 & \gamma_+\delta_2 & 0 & 0 & 0 & 0 & \eta(\beta_- - \epsilon_-) & -\eta(\beta_+ + \epsilon_+) & -\eta(\beta_- + \epsilon_+) & \eta(\beta_+ - \epsilon_-) \\
-\gamma_+\delta_2 & \gamma_-\delta_1 & 0 & 0 & 0 & 0 & \eta(\beta_+ + \epsilon_+) & -\eta(\beta_- - \epsilon_-) & -\eta(\beta_+ - \epsilon_-) & \eta(\beta_- + \epsilon_+) \\
0 & 0 & -\eta(\beta_- - \epsilon_+) & -\eta(\beta_+ + \epsilon_-) & -\eta(\beta_- - \epsilon_-) & -\eta(\beta_+ + \epsilon_+) & 0 & 0 & 0 & 0 \\
0 & 0 & \eta(\beta_+ + \epsilon_-) & \eta(\beta_- - \epsilon_+) & \eta(\beta_+ + \epsilon_+) & \eta(\beta_- - \epsilon_-) & 0 & 0 & 0 & 0 \\
0 & 0 & \eta(\beta_- + \epsilon_-) & \eta(\beta_+ - \epsilon_+) & \eta(\beta_- + \epsilon_+) & \eta(\beta_+ - \epsilon_-) & 0 & 0 & 0 & 0 \\
0 & 0 & -\eta(\beta_+ - \epsilon_+) & -\eta(\beta_- + \epsilon_-) & -\eta(\beta_+ - \epsilon_-) & -\eta(\beta_- + \epsilon_+) & 0 & 0 & 0 & 0 \\
\end{pmatrix},
\end{eqnarray}

where the rows and columns are labelled by $\ket{\Psi^{(0)}_+}, \ket{\Psi^{(0)}_-}, \ket{\Psi^{(1)}_+}, \ket{\Psi^{(1)}_-}, \ket{\widetilde{\Psi}^{(1)}_+}, \ket{\widetilde{\Psi}^{(1)}_-}, \ket{\Psi^{(2)}_+}, \ket{\Psi^{(2)}_-}, \ket{\widetilde{\Psi}^{(2)}_+}$ and $\ket{\widetilde{\Psi}^{(2)}_-}$ of (\ref{eig-spi-1}-\ref{eig-spi-2}), respectively. The coefficients $\beta_{\pm}$, $\gamma_{\pm}$, $\eta$, $\epsilon_{\pm}$ and $\delta_i$ for $i \in \{ 1,2,3,4 \}$ are given as follows

\begin{eqnarray}
\beta_{\pm} &=& \frac{\Lambda}{4} \left(\sqrt{2\kappa + 1} \pm \sqrt{\kappa + 1}\right)~,~
\gamma_{\pm} =  \left(1 \pm \sqrt{\kappa + 1} \right)
~,~ \epsilon_{\pm} = \frac{\Lambda}{4}\left( \sqrt{2\kappa}\sqrt{\kappa +1} \pm \sqrt{\kappa}\sqrt{2\kappa +1} \right),\\
\eta &=& \frac{X\sqrt{\kappa}}{\sqrt{(\kappa +1)(2\kappa +1)}}~,~ \delta_1 = \frac{X\Lambda}{2\sqrt{2(\kappa +1)}}\left( 1 +\sqrt{\kappa} + \sqrt{\kappa +1} \right) ~,~  \delta_2 = \frac{X\Lambda}{2\sqrt{2(\kappa +1)}}\left( 1 +\sqrt{\kappa} - \sqrt{\kappa +1} \right), \\
\delta_3 &=& \frac{X\Lambda}{2\sqrt{2(\kappa +1)}}\left( 1 -\sqrt{\kappa} + \sqrt{\kappa +1} \right)~,~\delta_4 = \frac{X\Lambda}{2\sqrt{2(\kappa +1)}}\left( 1 - \sqrt{\kappa} - \sqrt{\kappa +1} \right). \nonumber
\end{eqnarray}

Now the largest eigenvalue of the matrix corresponding to $M_l^\dagger M_l$ (using $Mathematica$) comes out to be $X^2\Lambda^2\kappa$. Therefore using the $C^*$ algebra property  $\Big\| M_l^\dagger M_l \Big\|_{op} = \Big\| M_l \Big\|_{op}^2$ the ball condition ($B_T$) becomes

\begin{equation} \label{ball-1}
\Big\| \left[\mathcal{D}_T,\mathcal{P}_2\pi\left( a_s^{(l)} \right) \mathcal{P}_2\right] \Big\|_{op} = X\Lambda\sqrt{\kappa} \leq 1,
\end{equation}  

which alongwith \eqref{tr_1} when substituted in \eqref{dis-1.2} and noting that $\kappa = \frac{2}{\theta\Lambda^2}$ reproduces the Moyal plane distance \eqref{dis-Moyal} as expected. Moreover as we increase the rank of the projection operator $\mathcal{P}_N$ from $N=2$ to $N=3$ and so on, we find that the ball condition \eqref{ball-1} remains unaffected and and so does the distance \eqref{dis-Moyal} for all order of $N$.
\\
As for the case of $c_1 = -c_2 = X$ we find, after following the same procedure, that the ball condition gets modified as 

\begin{equation}
\Big\| \left[\mathcal{D}_T,\mathcal{P}_2\pi\left(a_S^{(l)}\right)\mathcal{P}_2\right] \Big\|_{op} = X\Lambda\sqrt{8+\kappa+4\sqrt{1+\kappa}} \leq 1,
\end{equation} 

which together with \eqref{tr_1} and \eqref{dis-1.2} gives following estimate of the distance distance

\begin{equation}
d\left( \Omega^{(z)}_i,\Omega^{(0)}_i \right)_{est} = \frac{2\abs{z}}{\Lambda\sqrt{8+\kappa+4\sqrt{1+\kappa}}},
\end{equation}

which is clearly less than \eqref{dis-Moyal} and is therefore rejected.

\subsection{Transverse distance} \label{sec_trans}

For the transverse case we take as an ansatz for the optimal algebra element $a_s^{(t)} := \mathds{1}_{\mathcal{H}_q}\otimes a_2$ (the superscript `t' stands for transverse) such that it's representation looks like

\begin{equation}
\pi\left( a_s^{(t)} \right) = \begin{pmatrix}
\mathds{1}_{\mathcal{H}_q} & 0 \\ 0 & \mathds{1}_{\mathcal{H}_q}
\end{pmatrix} \otimes \begin{pmatrix}
c_1 & 0 \\ 0 & c_2
\end{pmatrix},
\end{equation}

then the Connes distance \eqref{Connes-dist} between states $\Omega^{(z)}_1$ and $\Omega^{(z)}_2$ corresponding to transverse distance $d_{PP'}$  (see Figure 3)

\begin{equation} \label{tr_2}
d_t\left( \Omega^{(z)}_1, \Omega^{(z)}_2 \right)  = \sup_{a_t \in B_T} \abs{\Tr_M\left( d\Omega^{(z)} a_t \right)} ~~;~~ d\Omega^{(z)} = \rho_z\otimes (\omega_1 - \omega_2)
\end{equation}

comes out to be same as that in two point space \eqref{trans_dis}. However the algebra of Moyal plane is non-unital and thus $\mathds{1}_{\mathcal{H}_q} \not\in \mathcal{A}_M =  \mathcal{H}_q$. In order to get around this problem we again make use of the projection operator $\mathds{P}_N$ \eqref{a13} as in \eqref{limit} and instead of $\pi\left( a_s^{(t)} \right)$ we work with

\begin{equation} \label{alg_trans}
\pi_N \left( a_s^{(t)} \right) =  \begin{pmatrix}
P_N & 0 \\ 0 & P_{N-1}
\end{pmatrix} \otimes \begin{pmatrix}
c_1 & 0 \\ 0 & c_2
\end{pmatrix}.
\end{equation}

With this we get the trace in \eqref{tr_2} which comes out to be

\begin{equation} \label{tr_3}
\Tr_M\left( d\Omega (z) a_s^{(t)} \right) = \frac{1}{2}\Tr_T\left( \pi\left( d\Omega (z) \right) \pi_N \left( a_s^{(t)} \right) \right) = c_1 - c_2
\end{equation}

where $\Tr_T$ denotes tracing over $\mathcal{H}_T$ and the extra factor of $\frac{1}{2}$ arises precisely to compensate the double counting, stemming from the diagonal representation \eqref{a9}. The matrix representation of $\left[\mathcal{D}_T,\pi_N \left( a_s^{(t)} \right) \right] =: M_t$ (say) comes out to be a block-diagonal matrix

\begin{equation} \label{mat3}
M_t = \begin{pmatrix}
Q & 0_{2\times 4} & \hdots & 0_{2\times 4} \\
0_{4\times 2} & R^{(1)} & \hdots & 0_{4\times 4} \\
\vdots & \vdots & \ddots & & \\
0_{4\times 2} & 0_{4\times 4} & & R^{(N)}
\end{pmatrix},
\end{equation}

where $0_{m\times n}$ are the $m \times n$ null rectangular matrices while the $2 \times 2$ square matrix $Q$ and $4 \times 4$ square matrices $R^{(m)}$ with $m \in \{ 1,2,3... \}$, being a positive integer in the range $1 \leq m \leq N$ are given by

\begin{equation} \label{mat4}
Q = \begin{pmatrix}
0 & \Lambda Y \\
-\Lambda Y & 0
\end{pmatrix} ~~;~~  R^{(m)} = \frac{\Lambda Y}{\sqrt{m\kappa +1}} \begin{pmatrix}
0 & -\sqrt{m\kappa} & 0 & 1 \\
\sqrt{m\kappa} & 0 & -1 & 0 \\
0 & 1 & 0 & \sqrt{m\kappa} \\
-1 & 0 & -\sqrt{m\kappa} & 0
\end{pmatrix}~;~ Y = c_1 - c_2.
\end{equation}

With this the matrix $M_t^\dagger M_t$ becomes proportional to the identity matrix
\begin{equation} \label{mat5}
M_t^\dagger M_t = \Lambda^2 Y^2 \begin{pmatrix}
\mathds{1}_{2 \times 2} & 0_{2\times 4} & \hdots & 0_{2\times 4} \\
0_{4\times 2} & \mathds{1}_{4 \times 4} & \hdots & 0_{4\times 4} \\
\vdots & \vdots & \ddots & & \\
0_{4\times 2} & 0_{4\times 4} & & \mathds{1}_{4 \times 4}\
\end{pmatrix}
\end{equation}

\begin{wrapfigure}{l}{8cm}
\includegraphics[width=8cm]{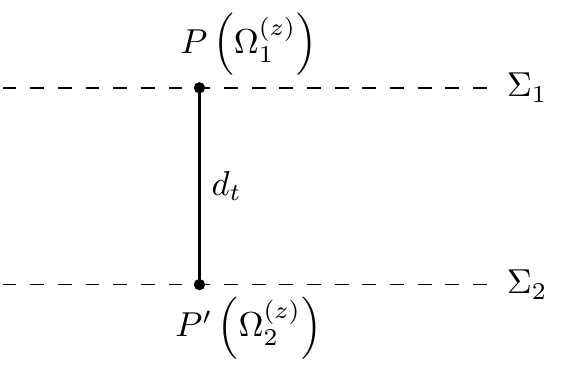}
\caption{Identical states belonging to different Moyal planes.}
\end{wrapfigure}

enabling us to just read-off the operator norm as $\| M_t^\dagger M_t \|_{op} = \abs{Y}^2 \Lambda^2$ which is independent of both $m$ and $N$ and so the ball condition ($B_T$) can be written for any arbitrary large $N$ as,

\begin{equation} \label{ball-2}
\Big\| \left[\mathcal{D}_T,\pi_N \left( a_s^{(t)} \right) \right] \Big\|_{op} = \abs{Y}\Lambda \leq 1,
\end{equation}

which together with \eqref{tr_3} when substituted in \eqref{tr_2} yields the transverse distance \eqref{trans_dis}. Finally in the limit $N \to \infty$, $\pi_N \left( a_s^{(t)} \right) \to \pi \left( a_s^{(t)} \right)$, we essentially recover the same result of \cite{b20}, obtained by unitizing the Moyal plane algebra.

\subsection{Hypotenuse distance}

To find the hypotenuse distance $d_{P'Q}$, we consider the states $\Omega^{(z)}_1$ and $\Omega^{(0)}_2$ as shown in Figure 4. We then proceed by making an ansatz about the optimal element of the algebra $\mathcal{A}_T$, $a_s^{(h)}$ (superscript `h' stands for hypotenuse), as a linear combination of the optimal elements of the longitudinal \eqref{A1} and transverse \eqref{alg_trans} optimal element so that it's representation takes following form

\begin{equation}\label{hypo-state}
\pi_N \left( a_s^{(h)} \right) = \begin{pmatrix}
b + b^\dagger & 0 \\ 0 & b + b^\dagger
\end{pmatrix}\otimes \begin{pmatrix}
X & 0 \\ 0 & X
\end{pmatrix}  + \begin{pmatrix}
P_N & 0 \\ 0 & P_{N-1}
\end{pmatrix} \otimes \begin{pmatrix}
c_1 & 0 \\ 0 & c_2
\end{pmatrix}.
\end{equation} 

Note that we have choosen $c_1 = c_2 = X$ case for the $\pi \left( a_l \right)$ as that is what gave us the Moyal plane distance. With this $\pi \left( a_s^{(h)} \right)$ we obtain, after a bit of calculation, following trace

\begin{equation}
\Tr_T\left( \pi_N \left( a_s^{(h)} \right) \pi \left( \Omega^{(z)}_1 - \Omega^{(0)}_2 \right) \right) = 4zX + 2c_1e^{-\abs{z}^2}\left( 1 + \abs{z}^2  + \hdots + \frac{\abs{z}^{2N}}{N!} \right)  -  2c_2 ~,
\end{equation}

which in the limit $N\rightarrow \infty$ takes the form

\begin{equation}
\Tr_T\left( \pi \left( a_s^{(h)} \right) \pi \left( \Omega^{(z)}_1 - \Omega^{(0)}_2 \right) \right) = 2\left( 2zX + Y \right),
\end{equation}

where $Y$ is same as in \eqref{mat4}. With this the Connes distance \eqref{Connes-dist} in this case takes following form.

\begin{eqnarray} \label{dis-2.1}
d_h\left( \Omega^{(z)}_1, \Omega^{(0)}_2 \right) &=& \sup_{a_s^{(h)} \in B_T} \abs{  Tr_M\left( \left( \Omega^{(z)}_1 - \Omega^{(0)}_2 \right) a_s^{(h)} \right) } \\ 
&=& \sup_{a_s^{(h)} \in B_T} \frac{1}{2} \abs{ Tr_T\left( \pi\left( \Omega^{(z)}_1 - \Omega^{(0)}_2 \right) \pi \left( a_s^{(h)} \right) \right) } \\ 
&=& \sup_{a_s^{(h)} \in B_T} \abs{ 2zX + Y} \label{dis-2.2}
\end{eqnarray} 

Now the matrix representation of $\left[\mathcal{D}_T,\mathcal{P}_N\pi_N (a_s^{(h)}) \mathcal{P}_N \right] =: M_h$ is just the sum of the matrices for longitudinal \eqref{mat2} and transverse \eqref{mat3} cases. 

\begin{wrapfigure}{l}{8cm}
\includegraphics[width=8cm]{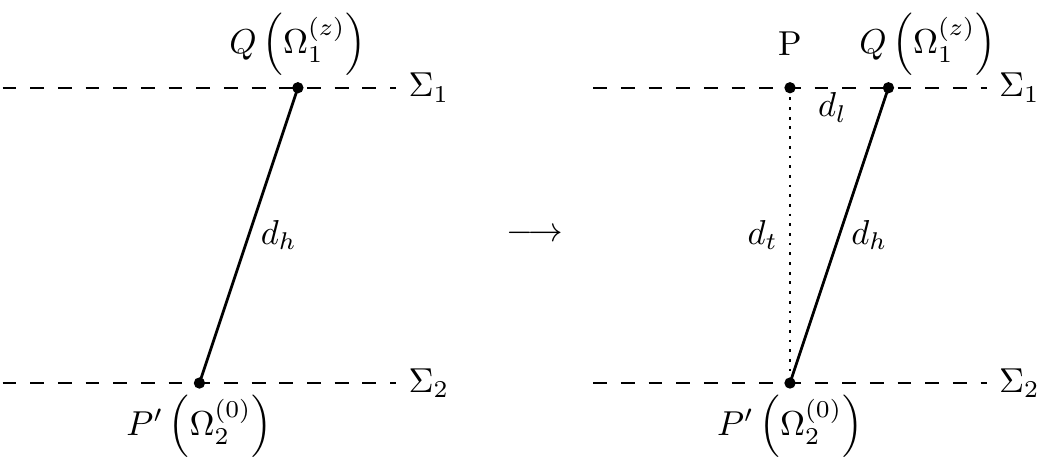}
\caption{Different states belonging to different Moyal planes.}
\end{wrapfigure}

The largest eigenvalue for the matrix $M_h^\dagger M_h$ comes out to be $\Lambda^2 \left( \kappa X^2 + Y^2 \right)$ and we get following ball condition ($B_T$)

\begin{equation} \label{ball-3} 
\Big\| \left[\mathcal{D}_T,\mathcal{P}_N\pi\left(a_s^{(h)}\right)\mathcal{P}_N\right] \Big\|_{op} = \Lambda\sqrt{\kappa X^2 + Y^2} \leq 1 
\end{equation}

In order to solve for the distance \eqref{dis-2.2} subject to the ball condition \eqref{ball-3} let's consider following lemma which can easily be proved (see e.g. Lemma 7 of \cite{Martinetti2}). \\

\textbf{Lemma}: For any $ \alpha, \beta \ge 0$,
\begin{equation}
 \sup_{{x^2 + y^2 \ \le \ 1}} \left(\alpha x + \beta y \right)= \sqrt{\alpha^2 + \beta^2 }
\end{equation}




Now, by appropriately identifying the symbols in the above lemma as $\alpha := \frac{2\abs{z}}{\sqrt{\kappa}\Lambda} = \sqrt{2\theta}\abs{z}, ~\beta := \frac{1}{\Lambda},~ x := \Lambda\sqrt{\kappa} X$ and $y := \Lambda Y$, we obtain following hypotenuse distance

\begin{equation}
d_h\left( \Omega^{(z)}_1, \Omega^{(0)}_2 \right) = \sqrt{2\theta\abs{z}^2 + \frac{1}{\abs{\Lambda}^2}}
\end{equation}
Now making use of (\ref{long_dis}, \ref{trans_dis}), this can be re-cast as,
\begin{equation}
d_h \left( \Omega^{(z)}_1, \Omega^{(0)}_2 \right) = \sqrt{\left( d_l\left( \Omega^{(z)}_i,\Omega^{(0)}_i \right) \right)^2 + \left( d_t\left( \Omega^{(0)}_1, \Omega^{(0)}_2 \right) \right)^2}
\end{equation}

which is exactly the Pythagoras equality (see Figure 4) that we discussed earlier \eqref{Pytha}.

\section{Higgs field from internal fluctuation of total Dirac operator} \label{sec6}

In this pen-ultimate section, we first provide a brief review of fluctuated Dirac operator and the emergence of ``Higgs field" in the framework of noncommutative geometry and it's impact on the metric. For this, we especially follow \cite{b16, Martinetti3}. The notion of Higgs field comes automatically, along with other gauge fields, in the framework of tensor product of spectral triples like the one which describes double Moyal plane \eqref{prod_triple}. The concept of charge conjugation operator in spectral triple $(\mathcal{A}, \mathcal{H}, \mathcal{D})$ is brought about by the so called real structure $J$ which is an anti-unitary operator on $\mathcal{H}$ satisfying $J^2 = \epsilon$, $J\mathcal{D} = \epsilon' \mathcal{D}J$ and $J\gamma = \epsilon^{\prime\prime} \gamma J$, where $\epsilon$, $\epsilon'$ and $\epsilon^{\prime\prime}$ are restricted to take values $\pm 1$ only.  The KO-dimension $n \in \mathds{Z}_8$ is then determined by the values of $\epsilon$, $\epsilon'$ and $\epsilon^{\prime\prime}$ according to a standard table (see  \cite{b16}).
To preserve the operator $J$ under a unitary transformation $u \in \mathcal{A}$, satisfying $u^*u = uu^* = 1$ of $\pi(\mathcal{A})$, as in section \ref{sec3}, one has to use $U := uJuJ^{-1}$, which fluctuates the Dirac operator as

\begin{equation} \label{fluc_dirac}
\mathcal{D}_A = U\mathcal{D}U^\dagger = \mathcal{D} + A + \epsilon'JAJ^{-1}~;~ A = \pi(a_i)[\mathcal{D},\pi(b_i)]~\textsl{with} ~ a_i, b_i \in \mathcal{A}.
\end{equation}

Here $A's$ are nothing but the Clifford algebra valued one-forms of the triple $(\mathcal{A}, \mathcal{H}, \mathcal{D}_A)$. We suppress the summation index `i' from now on for brevity. The last term $JAJ^{-1}$ in \eqref{fluc_dirac} plays no role in distance calculation as it commutes with $\pi(a)~\forall~ a \in \mathcal{A}$ as can be shown easily by employing the so called first order axiom and axiom of reality (see, for instance, Lemma 5 in \cite{Martinetti3}). This leaves the ball condition B \eqref{Connes-dist} unchanged and we therefore drop this term. Actually, the charge conjugation operator $J$ maps an algebra element $a \in \mathcal{A}$  to the opposite algebra element $a^o$ satisfying $(ab)^o = b^oa^o$, so that $\pi(a^o) := J\pi(a)^\dagger J^{-1}~\forall~a \in \mathcal{A}$ can act from the right on the elements of Hilbert space $\mathcal{H}$. This, however, is not allowed in the present construction of spectral triple \eqref{spec_trip} as here we have only a left module, whereas we need a bi-module where both left and right actions are defined. To this end, consider the following spectral 

\begin{equation} \label{spec_trip_new}
\mathcal{\widetilde{A}}_T := \mathcal{A}_T = \mathcal{H}_q \otimes M_2^d (\mathds{C}) ~;~ \mathcal{\widetilde{H}}_T = \left( \mathcal{H}_q \otimes M_2 (\mathds{C}) \right) \otimes M_2^d (\mathds{C}) \ni \widetilde{\Psi} ~;~ \mathcal{\widetilde{D}}_T \widetilde{\Psi} = \mathcal{D}_T \widetilde{\Psi} + \widetilde{\Psi} \mathcal{D}_T,
\end{equation}

instead of \eqref{spec_trip}. Note that we have just replaced $\mathcal{H}_T$ in \eqref{spec_trip} by $\widetilde{\mathcal{H}}_T$ here by replacing two factors of $\mathds{C}^2$ in $\mathcal{H}_T$ by $M_2 (\mathds{C})$ and $M^d_2 (\mathds{C})$ respectively. Here the representation $\pi$ is as before \eqref{piat} and grading operator   $\widetilde{\gamma}_T \widetilde{\Psi} = \gamma_T \widetilde{\Psi} + \widetilde{\Psi} \gamma_T$, with the $\gamma_T$ as in \eqref{gamma_total}. Note that there is no difference in the structures of $\mathcal{D}_T$ and $\widetilde{\mathcal{D}}_T$; the difference is given by their action on $\widetilde{\Psi} \in \widetilde{\mathcal{H}}_T$. More precisely, the action of latter is given by the sum of left and right actions of the former.  Likewise for $\gamma_T$ and $\widetilde{\gamma}_T$. On this spectral triple we can define charge conjugation operator $J_T$ as $J_T \widetilde{\Psi} = \widetilde{\Psi}^\dagger$ (i.e. Hermitian conjugate of $\widetilde{\Psi}$), which implies that $J_T^{-1} = J_T$ as $J_T^\dagger = J_T^{-1}$. With this real structure $J_T$ the right action $\mathcal{D}_T$ can be represented by $J_T\mathcal{D}_T J_T^{-1} \widetilde{\Psi} = J_T \left( \mathcal{D}_T \widetilde{\Psi}^\dagger \right) = \widetilde{\Psi} \mathcal{D}_T$ by using the fact that Dirac operator is Hermitian. One can now check easily that $\epsilon = \epsilon' = \epsilon^{\prime\prime} = 1$ and the KO-dimension of this triple comes out to be 0 modulo 8. We therefore have $\mathcal{\widetilde{D}}_T = \mathcal{D}_T + J_T\mathcal{D}_T J_T^{-1}$ which again enables us to easily verify that $[\mathcal{\widetilde{D}}_T, \pi(a_T) ] = [\mathcal{D}_T, \pi(a_T)]$ for all $a_T \in \mathcal{A}_T$. In light of this we see that the ball condition B \eqref{Connes-dist} remains unaffected with the new spectral triple \eqref{spec_trip_new} and gives the same distance as the algebra $\mathcal{\widetilde{A}}_T$ as that of $\mathcal{A}_T$, making the two triples equivalent as far as the distance is concerned. We can therefore revert back to the previous definition of spectral triple of double Moyal plane \eqref{spec_trip} except that the Dirac operator is now augmented by a ``Higgs" term:

\begin{equation} \label{fluc_dirac_op}
\mathcal{D}_A = \mathcal{D}_T + A.
\end{equation}

To understand the splitting of the one form $A$ \eqref{fluc_dirac} in case of double Moyal plane, we consider the algebra elements to be of separable form like $a_T := a\otimes a_2$ and $b_T := b \otimes b_2 \in \mathcal{A}_T$ as in section \ref{sec5} so that we get

\begin{equation} \label{one-form}
\pi(a_T) [ \mathcal{D}_T, \pi(b_T)] = \pi(a) [ \mathcal{D}_M, \pi(b)] \otimes a_2 b_2 + \sigma_3 \pi(a b)\otimes a_2 [ \mathcal{D}_2, b_2].
\end{equation}

Note that we have used the fact that chirality operator commutes with representation of algebra elements in the above calculation. The first term on the right hand side of \eqref{one-form} contains one form of Moyal plane i.e. $\pi(a) [ \mathcal{D}_M, \pi(b)]$, which gives rise to gauge field in non-commutative geometry. On the other hand the additional second term  on the RHS of \eqref{one-form} has the one form of two point space i.e. $a_2 [ \mathcal{D}_2, b_2]$, which gives rise to the prototype of the scalar Higgs field. For our discussion here we only focus on the implication of this ``Higgs field" on the metric aspect particularly on the transverse distance and thus retaining only this term the one-form $A$ becomes

\begin{equation} \label{one-form}
A = c \sigma_3 \otimes  a_2 [ \mathcal{D}_2, b_2] ; ~ c := a b \in  \mathcal{H}_q.
\end{equation}

It is noteworthy that we can write this as $A = \sigma_3 \otimes H$, where $H = c  a_2 [ \mathcal{D}_2, b_2]$ is referred to as the ``Higgs field" in the literature. This is, however, in our case just a prototype scalar Higgs field as we are dealing with 2 dimensional Moyal plane. Now in order to compute the transverse distance $d_t(\rho_z\otimes \omega_1, \rho_z\otimes \omega_2)$ between a general state $\rho_z = |z\rangle\langle z|$ in one Moyal plane and it's clone in the other one i.e. the counterpart of \eqref{trans_dis} in presence of ``Higgs" field, we begin by constructing the restricted spectral triple using the prescription \eqref{restricted_triple} with the projection operator $\mathcal{P}_{N=\tilde{0}}$ \eqref{proj-3}, which is same as $P^{(trans)}_{T(z)}$ \eqref{proj_translated}. But to ensure that this transverse distance can again be computed by this restricted triplet itself, we have to ensure that the condition \eqref{proj_cond} is satisfied here too i.e.

\begin{equation}
\Big[ \mathcal{D}_A, P^{(trans)}_{T(z)} \Big] = 0.
\end{equation}

By making use of (\ref{proj_translated}, \ref{one-form}) this essentially boils down to 

\begin{equation} \label{cond_new}
\big[ c, |\tilde{0}\rangle\langle \tilde{0}| \big] = \big[ c, |z\rangle\langle z| \big] 0,
\end{equation}

which can be easily shown to satisfy, for example, if $c$ belongs the the subspace $\mathcal{H}_q^{\tilde{0}} := Span\big\{ \ket{\tilde{m}}\bra{\tilde{n}} : m,n \in \{ 1,2,3...\} \big\}$. We shall, for simplicity, assume in the following that this condition holds. Otherwise, one has to clearly make use of complete spectral triple. Clearly if \eqref{cond_new} holds for a particular state $\rho_z = |z\rangle\langle z|$ does not ensure that it holds for other states $\rho_{z'} = |z'\rangle\langle z'|$ also. Now, any arbitrary algebra element $a_T \in \mathcal{A}_T$ \eqref{spec_trip} gets projected as (for $a \in \mathcal{A}_M$ and $c_1, c_2 \in \mathds{C}$)

\begin{equation} \label{alg_new}
\pi(a_T) = \begin{pmatrix}
a & 0 \\ 0 & a
\end{pmatrix} \otimes \begin{pmatrix}
c_1 & 0 \\ 0 & c_2
\end{pmatrix} \mapsto \mathcal{P}_{N=\tilde{0}} \pi(a_T) \mathcal{P}_{N=\tilde{0}} = \begin{pmatrix}
|\tilde{0}\rangle\langle \tilde{0}| & 0 \\ 0 & 0
\end{pmatrix} \otimes \begin{pmatrix}
fc_1 & 0 \\ 0 & fc_2
\end{pmatrix};~ f = \bra{\tilde{0}}a\ket{\tilde{0}} = \langle z|a|z\rangle.
\end{equation}

Also, using the algebra element $a_2 = \begin{pmatrix}
\alpha_1 & 0 \\ 0 & \alpha_2
\end{pmatrix},~ b_2  = \begin{pmatrix}
\beta_1 & 0 \\ 0 & \beta_2
\end{pmatrix} \in M_2^d (\mathds{C})$  with $\alpha_1, \alpha_2, \beta_1, \beta_2 \in \mathds{C}$ in \eqref{one-form} for the two-point space, the fluctuated Dirac operator \eqref{fluc_dirac_op} gets projected as

\begin{equation}
\mathcal{D}_A \mapsto \mathcal{P}_{N=\tilde{0}} \mathcal{D}_A \mathcal{P}_{N=\tilde{0}} = \begin{pmatrix}
|\tilde{0}\rangle\langle \tilde{0}| & 0 \\ 0 & 0
\end{pmatrix} \otimes \begin{pmatrix}
0 & \Lambda (1+g\alpha_1(\beta_2 - \beta_1)) \\ \bar{\Lambda} (1+g\alpha_2(\beta_1 - \beta_2)) & 0
\end{pmatrix},
\end{equation}

where $g = g(x_1, x_2) = \bra{\tilde{0}} c(\hat{x}_1, \hat{x}_2) \ket{\tilde{0}} = \langle z| c(\hat{x}_1, \hat{x}_2) |z\rangle$  is  some function of the dimensionful coordinates $x_1$ and $x_2$ \eqref{a5} and $\Lambda \in \mathds{C}$ (Note that the unitary transformation using $U(\phi)$ \eqref{unitary_op} to render $\Lambda$ real, as in Sec \ref{sec3}, has not been performed here). Further, the demand that the projected Dirac operator be Hermitian yields $\overline{g\alpha_1(\beta_2 - \beta_1)} = g\alpha_2(\beta_1 - \beta_2)$. Also, any arbitrary element of the Hilbert space $\Psi :=  \begin{pmatrix}
\ket{\psi_1} \\ \ket{\psi_2}
\end{pmatrix} \otimes \begin{pmatrix}
c_1 \\ c_2
\end{pmatrix} \in \mathcal{H}_T$ \eqref{spec_trip} gets projected as

\begin{equation}
\Psi \mapsto \mathcal{P}_{N=\tilde{0}} \Psi = \begin{pmatrix}
\ket{\tilde{0}} \\ 0
\end{pmatrix} \otimes \begin{pmatrix}
hc_1 & 0 \\ 0 & hc_2
\end{pmatrix}~;~ h = \langle \tilde{0}|\psi_1 \rangle ,
\end{equation}

so that the projected spectral triple takes following form

\begin{equation}
\mathcal{A}_T^{(\mathcal{P}_{N=\tilde{0}})} = \begin{pmatrix}
|\tilde{0}\rangle\langle \tilde{0}| & 0 \\ 0 & 0
\end{pmatrix} \otimes M_2^d(\mathds{C});~ \mathcal{H}_T^{(\mathcal{P}_{N=\tilde{0}})} = \begin{pmatrix}
\ket{\tilde{0}} \\ 0
\end{pmatrix} \otimes \mathds{C}^2;~ \mathcal{D}_T^{(\mathcal{P}_{N=\tilde{0}})} = \begin{pmatrix}
|\tilde{0}\rangle\langle \tilde{0}| & 0 \\ 0 & 0
\end{pmatrix} \otimes \begin{pmatrix}
0 & \Lambda(x_1, x_2) \\ \overline{\Lambda(x_1, x_2)} & 0
\end{pmatrix},
\end{equation}

where $\Lambda(x_1, x_2) = \Lambda (1 + \alpha_1 (\beta_2 - \beta_1)g(x_1, x_2) )$ is just another complex number. This has the same form as in \eqref{proj_triple} and is, therefore, has the structure of spectral triple of two point space. The transverse spectral distance i.e. the distance between the pure states $\rho_z$ and it's clone is identical to the spectral distance between the pair of states $\omega_1$ and $\omega_2$ of the restricted spectral triple and is given, in this case by

\begin{equation}
d_t(\rho_z\otimes \omega_1, \rho_z\otimes \omega_2) := d^{(rest)}_t(\omega_1, \omega_2) = \frac{1}{\abs{\Lambda(x_1, x_2)}}.
\end{equation}

This clearly fluctuates along the Moyal plane and reproduces \eqref{trans_dis} in the absence of ``Higgs" field. Thus this variation of transverse distance in presence/absence of all pervading ``Higgs" field provides an alternative geometrical perspective about ``Higgs" field itself.

\section{Conclusion}

Finite matrix spaces plays a crucial role in the formulation of ``standard model" in the framework of noncommutative geometry \cite{1st, b16}. On the other hand Moyal plane is one of the promising candidate for modelling geometry near Plank scale. The merger of these two i.e. double Moyal plane is therefore a very interesting toy model and we have explored it's metric structure using the Connes prescription of spectral distances in a Hilbert-Schmidt operatorial framework, which facilitates the construction of Dirac eigen-spinors and provides a natural basis to work with, which in turn helps us significantly in economising our calculations and reproducing several results existing in the literature \cite{Martinetti3, b20}. \\
The actual calculation of various distances is done using specific `optimal' algebra elements, which are partially motivated from our previous works \cite{b18}. The distances comes out very neatly and fits the Pythagorean equality exactly as expected. We did encounter the need to introduce the identity element of Moyal algebra to render it unital but, as an alternative approach to \cite{b20}, we worked with a sequence of projection operators constructed from Dirac eigen-spinors, which in the limiting case ($N \rightarrow \infty$) represents the identity element. \\
Finally, we fluctuate the Dirac operator and focus on the ``Higgs" field part and analyse the variation of transverse distance in the two point space. In order to fluctuate the Dirac operator we had to modify, at an intermediate stage, the spectral triple to arrive at an equivalent triple, so as to be able to incorporate the real structure. At the end, however, we reverted back to our earlier triple, except that the Dirac operator is now augmented with ``Higgs" term. As a consequence,  we found how the transverse distance, obtained by constructing a suitable projection operator to restrict the spectral triple to the two-point space depends on the coordinates of the Moyal plane. This projection operator, in turn, had to be necessarily constructed with these eigen-spinors, thus illustrating the important roles played by the eigen-spinors, which, we feel, were not emphasised adequately in the literature.

\section*{Acknowledgements} 

KK would like to acknowledge CSIR for financial support through JRF and also the authorities of SNBNCBS, Kolkata for kind hospitality and studentship where part of this research was carried out. The authors would also like to thank Dr. Sunandan Gangopadhyay and Aritra N. Bose for useful discussions at various stages.

\section*{Appendix} 

\subsection*{A: Spectral distance on Two point space}

Two-point space space is an abstract mathematical space of two complex numbers and the spectral triple for the same is given by (see \cite{b16}) 

\begin{equation} \label{2-point-triple}
\left( \mathcal{A}_2=\mathds{C}^2,\mathcal{H}_2=\mathds{C}^2,\mathcal{D}_2=\begin{pmatrix}
0&\Lambda\\
\bar{\Lambda}&0
\end{pmatrix} \right),
\end{equation}

where $\Lambda$ is a constant complex parameter of length-inverse dimension. Let's consider $a=\begin{pmatrix}
\lambda_1\\
\lambda_2
\end{pmatrix}=\lambda_1\begin{pmatrix}
1\\0
\end{pmatrix}+\lambda_2\begin{pmatrix}
0\\1
\end{pmatrix}\in \mathcal{A}_2=\mathds{C}^2$. Replacing the canonical bases $\begin{pmatrix}
1\\0
\end{pmatrix}$ and $\begin{pmatrix}
0\\1
\end{pmatrix}$ by the matrices $\begin{pmatrix}
1&0\\0&0
\end{pmatrix}$ and $\begin{pmatrix}
0&0\\0&1
\end{pmatrix}$, $\mathds{C}^2$ can now be thought of as ~Span$\Big\{\begin{pmatrix}
1&0\\0&0
\end{pmatrix},\begin{pmatrix}
0&0\\0&1
\end{pmatrix}\Big\}$, rather than ~Span$\Big\{\begin{pmatrix}
1\\0
\end{pmatrix},\begin{pmatrix}
0\\1
\end{pmatrix}\Big\}$ and $a\in\mathds{C}^2$ can be written as  
\begin{equation}
a=\begin{pmatrix}
\lambda_1&0\\0&\lambda_2
\end{pmatrix}\in  M_2^d(\mathds{C}), \label{a-in-C2}
\end{equation} 
where $ M_2^d(\mathds{C})$ is the c-valued $2\times 2$ diagonal matrix. In this construction we see that $\mathcal{A}_2$ has a manifest structure of an algebra by the usual matrix multiplication.

The action of $a\in\mathcal{A}_2$ on $\begin{pmatrix}
\mu_1\\ \mu_2
\end{pmatrix}\in\mathcal{H}_2=\mathds{C}^2$ is now given in following straightforward manner and we need not invoke any special representation $\pi$ here.

\begin{equation}
\pi(a)\begin{pmatrix}
\mu_1\\ \mu_2
\end{pmatrix}=a\begin{pmatrix}
\mu_1\\ \mu_2
\end{pmatrix}=\begin{pmatrix}
\lambda_1&0\\0&\lambda_2
\end{pmatrix}\begin{pmatrix}
\mu_1\\ \mu_2
\end{pmatrix}=\begin{pmatrix}
\lambda_1\mu_1\\\lambda_2 \mu_2
\end{pmatrix}. \label{a-action}
\end{equation} 

Then  a pure state $\omega_i (a)\in\mathds{C}$ on $\mathcal{A}_2$ is given by 
\begin{equation}
\omega_i (a)=\text{tr}(\omega_i \pi(a))=\text{tr}(\omega_i a) \label{2-p-state}
\end{equation}
so that it corresponds to the evaluation maps $\omega_1(a)=\lambda_1$ and $\omega_2(a)=\lambda_2$ on the two points of this space. Clearly, the choice 

\begin{equation} \label{2-pt-states} 
\omega_1=\begin{pmatrix}
1&0\\0&0
\end{pmatrix} ~;~ \omega_2=\begin{pmatrix}
0&0\\0&1
\end{pmatrix},
\end{equation}

 does the job, as can be seen easily by using (\ref{2-p-state}). Any other generic state is given by a convex sum of these two pure states and therefore are mixed in nature. These are clearly the density matrices associated to $\begin{pmatrix}
1\\0
\end{pmatrix}$ and $\begin{pmatrix}
0\\1
\end{pmatrix}$ and precisely corresponds to $2\times 2$ matrix basis, chosen for $\mathcal{A}_2=\mathds{C}^2$. Note that the dual space to $\mathds{C}^2$ is $\mathds{C}^2$ again, and as in the case of $a\in\mathcal{A}_2$, in (\ref{a-action}), the representation sign `$\pi$' becomes redundant. Now to compute the distance between these two points, let us consider a generic algebra element $a = \begin{pmatrix}
\alpha_1 & 0\\ 0 & \alpha_2
\end{pmatrix} \in \mathds{C}^2 \cong M_2^d(\mathds{C})$ and inserting it in the distance formula \eqref{Connes-dist} we obtain

\begin{equation} \label{2-pt-dis}
d(\omega_1, \omega_2) = \sup_{a \in B} \abs{\omega_1(a) - \omega_2(a)} = \sup_{a \in B} \abs{\alpha_1 - \alpha_2}.
\end{equation}

Moreover, the commutator $[\mathcal{D}_2,\pi(a)]$ takes the form

\begin{equation}
[\mathcal{D}_2,\pi(a)] = (\alpha_1 - \alpha_2)\begin{pmatrix}
0&-\Lambda\\
\bar{\Lambda}&0
\end{pmatrix},
\end{equation}

where we have used $\pi(a) = a$ in $\omega_1, \omega_2$ basis as before. Now invoking the $C^*$ property of the algebra $\mathcal{A}_2$ here i.e. $\| [\mathcal{D}_2, \pi(a)] \|_{op}^2 = \| [\mathcal{D}_2, \pi(a)]^\dagger[\mathcal{D}_2, \pi(a)] \|_{op}$ and noting that

\begin{equation}
[\mathcal{D}_2, \pi(a)]^\dagger[\mathcal{D}_2, \pi(a)] = \abs{\alpha_1 - \alpha_2}^2\abs{\Lambda}^2 \begin{pmatrix}
1 & 0 \\ 0 & 1
\end{pmatrix},
\end{equation}

we get $\| [\mathcal{D}_2, \pi(a)] \|_{op} = \abs{\alpha_1 - \alpha_2}\abs{\Lambda}$. With this the Lipschitz ball condition yields

\begin{equation} \label{2-pt-ball}
\abs{\alpha_1 - \alpha_2} \leq \frac{1}{\abs{\Lambda}}.
\end{equation}

Now applying supremum on \eqref{2-pt-ball} and substituting in \eqref{2-pt-dis} we get the spectral distance between pure states $\omega_1, \omega_2$ on the two point space as

\begin{equation} \label{2-pt-fin-dis}
d(\omega_1, \omega_2) = \frac{1}{\abs{\Lambda}}
\end{equation}
which is the known result of the spectral distance on the two-point space  \cite{b16}.

\end{document}